\documentstyle[12pt]{article}
\title{Generalization of the Coleman-Mandula Theorem \\
       to Higher Dimension}
\author{Oskar Pelc \and L. P. Horwitz}

\newcommand{\newsection}[1]{
 \vspace{10mm} \pagebreak[3]
 \addtocounter{section}{1}
 \setcounter{equation}{0}
 \setcounter{subsection}{0}
 \setcounter{paragraph}{0}
 \setcounter{equation}{0}
 \setcounter{figure}{0}
 \setcounter{table}{0}
 \addcontentsline{toc}{section}{\protect\numberline{\arabic{section}}{#1}}
 \begin{flushleft}
  {\large\bf \thesection. #1}
 \end{flushleft}
 \nopagebreak}

\renewcommand{\thesection}{\Roman{section}}

\newcommand{\qed}{\hfill \rule{2.5mm}{2.5mm}}
\newcommand{\al}{\alpha}
\newcommand{\A}{\mbox{\protect\boldmath$\cal A$}}
\newcommand{\Ag}{A_{g}}

\newcommand{\beq}{\begin{equation}}
\newcommand{\bbar}{\overline}
\newcommand{\B}{\mbox{\protect\boldmath$\cal B$}}
\newcommand{\Bcc}{\B_{c}}
\newcommand{\Bs}{\B_{s}}
\newcommand{\Bi}{\B^{\infty}}
\newcommand{\Bsm}{\B_{m}^{*}}
\newcommand{\dg}{^{\dagger}}
\newcommand{\der}[1]{\frac{d}{d#1}}
\newcommand{\derp}[1]{\frac{\partial}{\partial#1}}
\newcommand{\dl}{\delta}
\newcommand{\D}{{\cal D}}
\newcommand{\Dl}{\Delta}
\newcommand{\eeq}{\end{equation}}
\newcommand{\eg}{{\em e.g.\ }}

\newcommand{\f}{\varphi}
\newcommand{\fA}{f\cdot A}
\newcommand{\F}{\Phi}
\newcommand{\Fa}{\F_{\al}}
\newcommand{\FJ}{{\cal F}_{J}}
\newcommand{\Fn}{\F^{(n)}}

\newcommand{\Fns}{\Fn_{s}}

\newcommand{\Fone}{\F^{(1)}}
\newcommand{\Foned}{\F^{(1)'}}
\newcommand{\Fs}{\F_{s}}
\newcommand{\Ha}{\Hc_{\al}}
\newcommand{\Hc}{{\cal H}}
\newcommand{\Hn}{\Hc^{(n)}}
\newcommand{\Hns}{\Hn_{s}}
\newcommand{\Hone}{\Hc^{(1)}}
\newcommand{\Hs}{\Hc_{s}}
\newcommand{\ie}{{\em i.e.\ }}
\newcommand{\lm}{\lambda}
\newcommand{\inv}{^{-1}}
\newcommand{\La}{L^{\al}}
\newcommand{\Lc}{{\cal L}}
\newcommand{\Lm}{\Lambda}
\newcommand{\Lmp}{\Lm_{p}}
\newcommand{\Lp}{L^{(p)}}
\newcommand{\Lsq}{\Lc^{2}}
\newcommand{\m}{^{m}}
\newcommand{\ma}{m_{\al}}

\newcommand{\mm}{\mu_{m}}
\newcommand{\mma}{\mu_{\ma}}
\newcommand{\mn}{^{(m,n)}}
\newcommand{\ms}{^{(m)}}
\newcommand{\M}{{\cal M}}
\newcommand{\n}{^{n}}

\newcommand{\ns}{^{(n)}}
\newcommand{\Ns}{^{(N)}}
\newcommand{\one}{\mbox{\it1}}
\newcommand{\Om}{\Omega}
\newcommand{\pma}{p_{\ma}}

\newcommand{\psm}{p_{m}}
\newcommand{\pv}{\vec{p}}

\newcommand{\Pc}{{\cal P}}
\newcommand{\Pm}{P^{\mu}}

\newcommand{\s}{\sigma}
\newcommand{\sa}{\s_{\al}}
\newcommand{\so}{_{0}}
\newcommand{\sq}{^{2}}
\newcommand{\th}{\theta}
\newcommand{\tr}{\mbox{tr}}
\newcommand{\T}{\hat{\cal T}}
\newcommand{\TF}{\T_{F}}
\newcommand{\Tm}{\T_{m}}
\newcommand{\Tma}{\T_{\ma}}
\newcommand{\Ua}{U^{\al}}
\newcommand{\Uone}{U^{(1)}}
\newcommand{\Un}{U^{(n)}}
\newcommand{\UP}{{\cal U_{P}}}

\def\al{\alpha}

\def\dl{\delta}                \def\Dl{\Delta}

\def\lm{\lambda}               \def\Lm{\Lambda}

\def\th{\theta}               

               \def\Om{\Omega}
               
\def\Oc{\mbox{\protect$\cal O$}}
\def\Pc{\mbox{\protect$\cal P$}}

\def\Tc{\mbox{\protect$\cal T$}}

\def\Rb{{\bf R}}
\def\Zb{{\bf Z}}

\def\inv{^{-1}}

\def\hsc{\hspace{5mm},\hspace{5mm}}
\def\ie{{\em i.e.\ }}
\def\eg{{\em e.g.\ }}

\def\beq{\begin{equation}}
\def\eeq{\end{equation}}

\begin{document}


\begin{titlepage}

\begin{flushright}
RI-2-96 \\
TAUP 2175-94\\
IASSNS-96/31 \\
\today
\\[15mm]
\end{flushright}

\begin{center}
\Large
Generalization of the Coleman-Mandula Theorem \\
to Higher Dimension
\\[10mm]
\large
Oskar Pelc$^*$\footnote{E-mail: oskar@shum.cc.huji.ac.il}
\normalsize and \large 
L. P. Horwitz$\dg$\footnote{
    E-mail: horwitz@sns.ias.edu

 On sabbatical leave from the School of Physics and Astronomy, 
 Raymond and Beverly Sackler Faculty of Exact Sciences,
 Tel-Aviv University, Ramat-Aviv, Israel;  also at Department of Physics, 
 Bar-Ilan University, Ramat Gan, Israel.}
\normalsize \\[5mm]
{\em $^*$Racah Institute of Physics, The Hebrew University\\
  Jerusalem, 91904, Israel}
\normalsize \\[5mm]
{\em $\dg$School of Natural Sciences, Institute for Advanced Study,\\
 Princeton, N.J. 08540, USA}
\\[15mm]
\end{center}

\begin{abstract}
The Coleman-Mandula theorem, which states that space-time and internal
symmetries cannot be combined in any but a trivial way, is generalized
to an arbitrarily higher spacelike dimension. Prospects for
further generalizations of the theorem (space-like representations,
larger time-like dimension, infinite number of particle types) are
also discussed. The original proof relied heavily on the Dirac
formalism, which was not well defined mathematically at that time.
The proof given here is based on the rigorous version of the Dirac
formalism, based on the theory of distributions. This work serves also 
to demonstrate the suitability of this formalism for practical applications.
\end{abstract}

------------------------------


\end{titlepage}

\pagestyle{plain}
\setcounter{page}{1}
\flushbottom

\newsection{Introduction}

\subsection{The Coleman Mandula theorem}

Symmetry plays a key role in modern physics, and in the investigation
of the foundations of physics in particular. Symmetry considerations
were found extremely useful in the understanding of physical phenomena
({\em e.g.\ }particle classification, selection rules) and in the
formulation of theories describing a given physical system. The choice
of a symmetry group of the system determines to a great extent its
properties.

In a relativistic theory, this group must contain (as a subgroup)
the Poincar\'{e} group:
translations, rotations and Lorentz transformations.
In 1967, Coleman and Mandula \cite{Col-Man} proved a theorem which
puts a severe restriction on the groups that can serve as physical
symmetry groups.

They proved that (this is a loose statement of the theorem; a more
precise one will follow): if
\begin{quote}
\begin{enumerate}
  \item the S matrix is not trivial and is such that the elastic
        scattering amplitudes are analytic functions of $s$ (the
        squared center-of-mass energy) and $t$ (the squared momentum
        transfer),
  \item the mass spectrum of the one particle states is a (possibly
        infinite) set of isolated points, all positive, and there is
        a finite number of particle types with a given mass,
  \item G is a connected symmetry group of the S matrix which contains
        the Poincar\'{e} group and is generated, at least locally, by
        generators (``infinitesimal symmetry transformations'')
        representable as integral operators in ``momentum space'' with
        distributions as kernels,
\end{enumerate}
then
\begin{quote}
  G is locally isomorphic to the direct product of the Poincar\'{e}
  group and an internal symmetry group (``internal'' means symmetries
  that commute with the Poincar\'{e} group).
\end{quote}
\end{quote}

The implications of the theorem are far reaching. It implies that, at
least in the domain of classical groups, and under the stated
assumptions of the theorem, there is no connection between the
space-time symmetries and the other symmetries - those ``mixing''
different particle ``types'' ({\em e.g.\ }charge, flavour, color). This
means that the properties and structure of the Poincar\'{e} group are
of no help in choosing the set of other symmetries. But the most
important implication is that symmetries cannot relate particles with
different mass and spin and thus the hope to describe the full variety
of particle types through symmetry considerations was destroyed.

The possibility of supersymmetry was not envisaged by Coleman and
Mandula and the introduction of supersymmetries in 1973 (by Volkov and
Akulov \cite{Vol-Aku} and independently by Wess and Zumino
\cite{Wes-Zum}) offered a bypass of their theorem.
However, most of the ideas in the theorem apply to
supersymmetries as well and in 1974 they were exploited by Haag,
\L{}opusza\'{n}ski and Sohnius \cite{Haag} to put severe restrictions
on the possible supersymmetries.

The set of supersymmetries constitutes a ``Graded Lie Algebra'' (GLA)
which is a generalization of a Lie algebra and thus the
supersymmetries are interpreted as infinitesimal symmetry
transformations. A GLA $\cal A$ is a direct sum
of two vector spaces: The space ${\cal A}_{0}$ of even  (``bosonic'')
elements, related by commutators, and the space ${\cal A}_{1}$ of odd
(``fermionic'') elements, related by anticommutators. (${\cal A}_{0}$
is an ordinary Lie algebra so the Coleman-Mandula theorem applies to it
directly)

Haag {\em et al.\ }proved that (this is also a loose statement): if
\begin{quote}
\begin{enumerate}
  \item assumptions 1 and 2 of the Coleman-Mandula theorem are satisfied,
  \item the elements of $\cal A$ are (infinitesimal) symmetries of the
        S-matrix,
\end{enumerate}

then
\begin{enumerate}
  \item the bosonic generators, except those of the Poincar\'{e} group
        are all translation invariant Lorentz scalars ({\em i.e.\ },
        generators of internal symmetries),
  \item the fermionic generators are translation invariant Majorana
        spinors,
  \item the commutators and anticommutators are determined to a large
        extent (details will not be given here); in particular, if
        there are no internal symmetries, they are determined uniquely.
\end{enumerate}
\end{quote}

\subsection{Extension of the Theorem}

We discuss in the following some motivations for extending the
Coleman-Mandula theorem to higher dimensions.

Manifestly covariant relativistic quantum mechanics has been
formulated as a theory of the evolution of events in space-time
\cite{Time,mult-tau}. The wave functions are square integrable
functions on the four dimensional space-time. The principle strategy
in developing the theory is to take the Schr\"{o}dinger formulation of
non-relativistic quantum mechanics in three dimensional space and
generalize it, with appropriate interpretations, to the
relativistic space-time of events. In particular, in \cite{Time}, the
wave functions are parametrized by a universal world time $\tau$ which
is the analogue of the non-relativistic time as an evolution parameter,
and evolution in $\tau$ is generated by a Hamiltonian-like operator
$K$, which is a function of space and time coordinates and their
canonical conjugate coordinates - momentum and energy. (This must be a
Lorentz-invariant function, in order to have the manifest relativistic
covariance of dynamical evolution.) For example, for a spinless
particle, one may take
  \[  K=\frac{p_{\mu}p^{\mu}}{2M}+V(\rho)  \]
(with the signature $-+++$) where $V(\rho)$ is a real function of
  $  \rho \equiv x_{\mu}x^{\mu} = x^{2} - t^{2}  $
and M is a parameter (with the dimension of mass).

It is important to notice that here $\{p_{\mu}\}$ are four independent
variables, so the mass $m\equiv\sqrt{p_{\mu}p^{\mu}}$ is a dynamical
variable and not a fixed parameter. It is obviously connected with the
fact that space and time coordinates are all independent dynamical
variables \cite{Time}, so it is an intrinsic property of every theory
of this kind \cite{mult-tau}.

Recalling the Coleman-Mandula theorem, we see that it is not valid here
because it assumes a discrete mass spectrum. Indeed, in the formulation
of a theory of electromagnetic interactions \cite{Time}, one gets,
through the requirement of gauge invariance, (in the same way that classical 
electromagnetism introduced the Lorentz symmetry), a 5-D symmetry group:
${\cal P}(4,1)$ or ${\cal P}(3,2)$ (where ${\cal P}(r,s)$ denotes 
the inhomogeneous pseudo orthogonal group of signature $(r,s)$)
and this is not a direct product of the Poincar\'{e} group ${\cal P}(3,1)$ 
and another group \cite{ext-sym}.
So the Coleman-Mandula theorem and its extension to supersymmetry are
not available here to guide us in the choice of the symmetries of a
system \cite{Witten}.

However, a further analysis suggests that Coleman-Mandula's assumptions can 
be generalized to this framework in accordance with all the other changes. 
The assumptions about the mass spectrum were actually assumptions about the
number of particle types, since a particle type was identified with an
irreducible representation of the Poincar\'{e} group in the Hilbert
space of one-particle states and such a representation has a definite
mass. But in the higher dimensional theory, a change in mass is not 
interpreted as a
change of particle type. For example, in the free ($V=0$) spinless
theory, $m\equiv\sqrt{p_{\mu}p^{\mu}}$ is a constant of the motion and
with an appropriate calibration it can be put equal to $M$. It is
therefore
possible to interpret $M$ (or some function of $M$) as the intrinsic
``mass'' of the particle described, while interactions take the
particle off ``mass shell''.

Thus, a particle type cannot be identified here with an irreducible
representation of ${\cal P}(3,1)$. But if we have a higher-dimensional
symmetry group, as suggested by the generalized form of electromagnetism, 
an irreducible
representation of this group may be an adequate candidate to be
identified with a particle type. So it seems that in the theorem that
should be the analogue of the Coleman-Mandula theorem here, the Poincar\'{e}
group should be replaced by the larger symmetry group and the mass
$m\equiv\sqrt{p_{\mu}p^{\mu}}$ should be replaced by the analogue
Casimir operator in the larger group. This is the type of theorem that
will be proved in this work. In this theorem, the $\Pc(3,1)$ group will be
replaced by $\Pc(r,1)$. A further generalization to $\Pc(r,s)$ would also be 
desireble, however in the consideration of more than one time-like direction, 
there are several complications and it is not clear if such a generalization 
is possible. This issue is discussed in subsection 4.1. 

This work may also be relevant to the recent developments of Kaluza-Klein 
type theories for the construction of grand unified or string theories, where
one deals with generalized space-time manifold of many dimensions (with 
signature {\em e.g.\ }$(d-1,1)$). However, in applying the theorem proved 
here to these cases, the meaning of the assumptions on the spectrum of the 
Casimir operator introduced above must be carefully considered.

There exist generalizations of the Coleman-Mandula theorem in other 
directions: candidates for symmetry generators defined by conserved currents
are studied in \cite{Am-Re}; more general generators and non-local charges 
are considered in \cite{Bu-Lo-Ra}; Galilean field theories are
studied in \cite{Garber} and massless particles are considered in 
\cite{Strube}.

\subsection{The Mathematical Formulation}

The proof of the original Coleman-Mandula theorem relies heavily on the
``bra'' and ``ket'' formalism of Dirac, using a basis of ``plane-wave
states'', matrix elements of operators in this basis, $\delta$-functions
etc.\ . The original formalism, due to Dirac \cite{Dirac} (confined to a
Hilbert space) is not well defined and thus is not suitable for a formal
mathematical proof. Here we use a rigorous version of this formalizm
\cite{RHS}, based on replacing the traditional Hilbert space
by a Gel'fand triple. In \cite{Horwitz-Pelc} this approach was used to
construct the basis of plane-wave states in a (well-defined) generalized
sense and to give a precise meaning to other concepts used in this context,
such as generators of symmetry and scattering amplitudes. The results of
\cite{Horwitz-Pelc} are extensively used throughout this work and they are
summarized in appendix A.

\subsection{The Structure of This Work}

We begin, in section 2, with a concise description of the main concepts
needed for the formulation and proof of the theorem -- the space of states,
the S-matrix, symmetries etc.\ . Then a precise statement of the theorem
is given. In section 3 the theorem is proved along the lines described in
\cite{Col-Man}. In section 4 we comment on the theorem proved and on
possible further extensions. Appendix A collects the notation and definition
of all the concepts used in the text and in appendix B some technical
properties of two-body scattering are derived.

\newsection{The Theorem}

\subsection{The Scenario}

In relativistic scattering theory, the space $\Hs$ of physical
states is a direct sum of (complex, separable)
Hilbert spaces
\begin{equation}
  \Hs=\bigoplus_{n=0}^{\infty}\Hns
\end{equation}
where $\Hns$ is the space of $n$-particle states (thus
called ``$n$-particle space'') and is (isomorphic to) a closed subspace
of the completed tensor product of $n$ one-particle spaces:
\begin{equation}\label{Hns}
  \Hns\subset\Hn=\bbar{\bigotimes_{1}^{n}\Hone}.
\end{equation}
The elements of $\Hns$ are those elements of $\Hn$ which have the right
symmetry properties with respect to the exchange of identical
particles. This symmetry is not relevant in the subsequent analysis,
therefore we consider the larger space
\beq \Hc=\bigoplus_{n=0}^{\infty}\Hn. \eeq

According to the rigorized Dirac formalism \cite{RHS}, $\Hc$
belongs to a {\em Gel'fand triple} $(\Phi,\Hc,\Phi')$ of
topological vector spaces, where $\Phi$ is embedded in $\Hc$
continuously as a dense subspace and $\Phi'$ is the (strong) dual of
$\Phi$. This also induces a corresponding Gel'fand triple for any
closed subspace of $\Hc$

The S-matrix $S$ is assumed to be a unitary operator on $\Hs$. It can
be identified as an element of $L^{\times}(\F;\F')$ -- a continuous
antilinear map from $\F$ to $\F'$ -- and as such, it has a
corresponding kernel $<\!S\!>$ (more precisely -- kernels
$<\!S\mn\!>\in\D(\Om\m\times\Om\n)$ ).

A symmetry transformation of the S-matrix is defined to be a unitary
or antiunitary operator $U$ in $\Hs$ which satisfies:
\begin{enumerate}
  \item $\Hone$ is $U$-invariant, {\em i.e.\ }$U$ turns one-particle
    states into one-particle states.
  \item $U$ acts on many-particle states in accordance to their
    relation to the tensor product of one-particle states:
    \begin{equation}
      U(f_{1}\otimes\cdots\otimes f_{n})=
        (Uf_{1})\otimes\cdots\otimes(Uf_{n})
    \end{equation}
    (and thus, according to property 1, $\forall n,\;\Hns$ is
    $U$-invariant).
  \item $U$ commutes with $S$.
\end{enumerate}

Under the assumptions of the theorem, including translation invariance
of $S$, it can be shown that its kernel $<\!S\!>$ has the following form
\beq
  <\!S\mn\!>=<\!I\mn\!>
    -i(2\pi)^{d}\dl^{d}(\sum_{1}\m q_{j}-\sum_{1}\n p_{i})<\!T\mn\!>
\eeq
where $d$ is the dimension of the momentum space, $-i(2\pi)^{d}$
is a conventional normalization factor and $<\!T\!>$ is a generalized
function of the momenta $q_{j},p_{i}$, restricted to the domain defined
by the constraint
\beq  \sum_{1}\m q_{j}-\sum_{1}\n p_{i}=0.  \eeq
The values of $<\!T\!>$ are called ``scattering amplitudes''.

\subsection{The Statement of the Theorem}

{\bf The assumptions:}

(See remarks in section 4)
\begin{enumerate}
  \item $G$ is a connected group of symmetries of the
    S-matrix $S$.
  \item (Lorentz invariance) $G$ contains a subgroup
    $\Pc'\so$, locally isomorphic to $\Pc(r,1)$, the inhomogeneous
    pseudo orthogonal group of Lorentzian signature ($r,1$), where
    $r\geq3$.
  \item All particle types correspond to positive-energy
    time-like representations of $\Pc$, the universal covering group
    of $\Pc'\so$ (\ie the spectrum $\TF$ of the momentum operator $P$
    in the space $\Hone$ of one-particle states is contained in the
    forward light cone:
    \[  \forall p\in\TF \hsc p_{\mu}p^{\mu}>0 \hsc E\equiv p^{0}>0).  \]
  \item (Particle finiteness) The number of particle
    types is finite.
  \item (Existence of generators) $G$ is generated, at least locally, by
    generators represented in the one-particle space $\Hone$ by
    (generalized) integral operators in momentum space, with distributions
    as kernels (\ie there is a neighborhood of the identity in $G$, such
    that each element in that neighborhood belongs to a one parameter
    subgroup $g(t)$ and there exists $\Ag\in L^{\times}(\Fone;\Foned)$
    that satisfies:
    \beq
      (\psi|\Ag\f\!>=\frac{1}{i}\der{t}(\psi,\Uone(g(t))\f)|_{t=0} \hsc
      \forall\f,\psi\in\Fone,
    \eeq
    where $\Uone$ is the representation of $G$ in $\Hone$).
  \item (Analyticity) The scattering amplitudes $<\!T\!>$ are regular
    functions of the momenta.

    The amplitudes $<\!T^{(2,2)}\!>$ for scattering between two-particle
    states
    \[  (p_{1},p_{2})\longrightarrow(p'_{1},p'_{2})  \]
    are analytic functions of (see appendix B.1)
    \[  s=(p_{1}+p_{2})\sq\mbox{ and }t=(p'_{1}-p_{1})\sq  \]
    in some neighborhood of the physical region.
  \item (The occurrence of elastic scattering) The amplitudes for elastic
    scattering of two particles do not vanish identically.
\end{enumerate}
{\bf The result obtained:}
\begin{quote}
  $G$ is locally isomorphic to the direct product of $\Pc'\so$ and an
  internal symmetry group.
\end{quote}

\newsection{The Proof}

The proof of the theorem is divided naturally to three parts. Each
part is built as a sequence of lemmas and ends with a proposition
which states the final results of that part. The first two parts are
completely independent and the last part uses only the propositions
of the preceding parts.

The proof is based on an analysis of the generators of $G$, concluding
that these are always sums of generators of $\Pc$ and generators of
internal symmetry transformations. The analysis will be performed on
a space $\A$ of operators that includes the set of generators as a
subspace (but {\em a priori} may be larger). In appendix B.2
it is shown that, $\Phi^{(2)}$ is $S^{(2,2)}$-invariant. This allows the
following definition:
\begin{quote}
  \bf Definition: \em
  $\A$ is the set $\{A\}$ of elements of $L^{\times}(\Fone;\Foned)$
  for which $A\dg$ ``commutes'' with the S-matrix $S$ in
  $\Fone\times\Fone$ in the sense:
  \beq
    (S^{*}\psi|A\dg\f>=<A\psi|S\f),\;
    \forall\f,\psi\in\Fone\times\Fone
  \eeq
  where
  \beq
    A\f:=(A\otimes I+I\otimes A)\f=
    (A\f_{1})\otimes\f_{2}+\f_{1}\otimes(A\f_{2}).
  \eeq
\end{quote}
The properties of $\A$:
\begin{enumerate}
  \item Each generator of $G$ is a (self adjoint) element of $\A$
    (see appendix \ref{Gen}).
  \item $\A$ is a vector space and it is a closed subspace of
    $\D'(\Om\times\Om)$: each sum or integral of elements of $\A$
    which converges in $\D'(\Om\times\Om)$ is in $\A$.
  \item If $A\in\A$ and $(\Lm,a)\in\Pc$ then
    $\Uone(\Lm,a)'A\Uone(\Lm,a)\in\A$ (since $\Fone$ is
    $U(\Pc)$-invariant and $S$ commutes with $U(\Pc)$ ).
  \item If $A,B\in\A$, $\Fone$ is invariant under $A$ and $B$ and
    $AB,BA\in L^{\times}(\Fone;\Foned)$ then $[A,B]\in\A$. (Notice
    that $\A$ is not necessarily a Lie algebra: for \mbox{$A,B\in\A$},
    $AB$ or $BA$ may be undefined.)
\end{enumerate}
 
\subsection*{Part 1}

This part analyzes the dependence of elements of $\A$ on the momentum.
Let $A$ be an element of $\A$ and $f$ a test function on the momentum
space (an element of $\D(\T)$). We define
\beq\label{fA-def}
  \fA:=\int_{\cal T}daU(1,a)'AU(1,a))\tilde{f}(a)
\eeq
where $\tilde{f}$ is the Fourier transform of $f$:
\beq
  \tilde{f}(a):=\int_{\T}\frac{dp}{(2\pi)^{d}}f(p)e^{ip\cdot a}.
\eeq
{\bf Lemma 1:}
\begin{quote}\em
  $\fA$ is in $\A$ and its matrix elements are:
  \beq\label{fAel}  <p'|\fA|p>=f(p'-p)<p'|A|p>  \eeq
  which implies that
  \beq\label{fAsupp}
    \mbox{supp}[(\fA)\f]\subset\mbox{supp}(\f)+\mbox{supp}(f).
  \eeq
\end{quote}
Proof:
\begin{quote}
  For each $\f,\psi\in\Fone$
  \begin{eqnarray*}
    \lefteqn{(\psi|(\fA)\f>=}  \\
    & = & \int_{\cal T}da\tilde{f}(a)(\psi|U(1,a)'AU(1,a)\f>= \\
    & = & \int_{\cal T}da\int_{\T}\frac{dq}{(2\pi)^{d}}
          f(q)e^{iq\cdot a} \\
    &   & \;\;\int d\mu(p')d\mu(p)e^{-ip'\cdot a}
          (\psi|p'><p'|A|p><p|\f)e^{ip\cdot a}  \\
    & = & \int d\mu(p')d\mu(p)(\psi|p'><p'|A|p><p|\f)  \\
    &   & \;\;\int_{\cal T}da\int_{\T}\frac{dq}{(2\pi)^{d}}f(q)
          e^{i[q-(p'-p)]\cdot a}  \\
    & = & \int d\mu(p')d\mu(p)(\psi|p'><p'|A|p><p|\f)f(p'-p)
  \end{eqnarray*}
  where the second to last equality is due to Fubini's theorem for
  distributions. The last stage identifies $<\fA>$ as a distribution on
  $\Om\times\Om$ (an element of $\D'(\Om\times\Om)$ ), thus $\fA$ is
  in $\A$ (according to properties 2 and 3 of $\A$).

  Now if $p'\in\mbox{supp}[(\fA)\f]$ then there exists
  $p\in\mbox{supp}(\f)$ for which \linebreak $<p'|\fA|p>\neq 0$. Using
  (\ref{fAel}), this implies that $f(p'-p)\neq0$, which means that
  $\Dl p\equiv p'-p$ is in supp$(f)$, so for each
  $p'\in\mbox{supp}[(\fA)\f]$ we have
  \[  p'=p+\Dl p,\mbox{ where }p\in\mbox{supp}(\f)
      \mbox{ and }\Dl p\in\mbox{supp}(f).              \]
  \qed
\end{quote}
Let
\begin{quote}
  $0\neq\Dl p\so\in\T,\;\rho>0,\;\Dl R$ a ball of radius $\rho$
  around $\Dl p\so$ and $f$ a test function on $\T$, with support
  contained in $\Dl R$.
\end{quote}
It will be shown that in such a situation, for $\rho$ sufficiently
small, $\fA$ vanishes because $\fA\neq 0$ would contradict the
occurrence of elastic scattering (\mbox{assumption 7} of the theorem).
To show this, we will construct regions in $\T$ such that states with
support in these regions are not connected by the S-matrix. For this
it is needed to analyze the action of $\fA$ in momentum space. The
momentum support of elements of $\Fone$ (and thus also of elements of
$\Foned$) is contained in ``the physical region'' $\TF$, which is a
(finite) union of orbits $\{\Tm\}_{m\in\M}$ (each orbit being an $r$
dimensional hyperboloid, see \mbox{figure \ref{fig1})}.
\begin{figure}
  \vspace{7.5cm}
  \caption{The physical region $\TF$}
  \label{fig1}
\end{figure}
According to eq. (\ref{fAsupp}), to have $(\fA)\f\neq 0$ for some
$\f\in\Fone$, supp($\f$) must contain $p\in\TF$ for which there exists
$\Dl p\in\Dl R$ that satisfies $p+\Dl p\in\TF$. In other words,
$p+\Dl R$ must intersect $\TF$ non-trivially. The set of all points in
$\Tm$ satisfying this condition is
\beq  R_{m}:=\Tm\cap(\TF-\Dl R).  \eeq
We  have therefore shown that
\beq  R:=\bigcup_{m\in\M}R_{m}  \eeq
obeys:

\noindent{\bf Lemma 2:}
\begin{quote}\em
  If $\f\in\Fone$ is such that $R\cap\mbox{\em supp}(\f)=\emptyset$
  then $(\fA)\f=0$.
\end{quote}
This lemma will be used to construct states annihilated by $\fA$. To
do this, we must describe $R$:
\beq  R_{m}=\Tm\cap(\TF-\Dl p\so+(\Dl p\so-\Dl R))  \eeq
and $\Dl p\so-\Dl R$ is a $\rho$-neighborhood of the origin so
$\TF+(\Dl p\so-\Dl R)$ is a $\rho$-neighborhood of $\TF$ and thus
$R_{m}$ is a neighborhood of $\Tm\cap(\TF-\Dl p\so)$ in $\Tm$. The
situation is described in figure \ref{fig2}. This is a cross section of
the momentum space at the plane $(p_{x},E)$ (where the $x$ axis was
chosen in the direction of $\Dl\pv\so$). Each shaded area is (the cross
section of) $\Tma-\Dl R$ (for some representative orbit $\Tma$) which
is identified as a $\rho$-neighborhood of $\Tma-\Dl p\so$ (denoted by
a dashed line). In this cross section, $R_{m}$ is a union of intervals,
the intersection of the hyperbola $\Tm$ with the shaded areas.
Since this is a cross section, each interval represents a band in the
full momentum space. These bands are described in figure \ref{fig3}.
This is a projection of $\Tm$ on the hyperplane orthogonal to the
$E$-axis (since the situation is rotationally symmetric around
$\Dl\pv\so$, $p_{y}$ can represent all the space-like directions
orthogonal to $p_{x}$).
\begin{figure}
  \vspace{7.0cm}
  \caption
    {A cross section of the momentum space at the plane $(p_{x},E)$}
  \label{fig2}
  \caption{A projection of $\Tm$ on the $\pv$-hyperplane}
  \label{fig3}
  \vspace{7.0cm}
\end{figure}

Explicitly, $\Tm\cap(\Tma-\Dl p)$ is the solution of
\beq
  E\sq=m\sq+\pv\,\sq\;;\;(E+\Dl E)\sq=\ma\sq+(\pv+\Dl\pv\,)\sq.
\eeq
In the $(p_{x},p_{y})$ plane it is characterized by the equation
\begin{eqnarray}
  \lefteqn{\Dl E\sq p_{y}\sq+\Dl m\sq p_{x}\sq+
           (\Dl m\sq+m\sq-\ma\sq)\Dl pp_{x}=}  \\
  & & \frac{1}{4}(\Dl m\sq+m\sq-\ma\sq)\sq-\Dl E\sq m\sq  \nonumber
\end{eqnarray}
(where $\Dl m\sq:=\Dl E\sq-\Dl p\sq,\;\Dl p:=||\Dl\pv\,||=\Dl p_{x}$)
\begin{itemize}
  \item for $\Dl m\sq=0$:
    \beq
      \Dl p\sq(p_{y}\sq+m\sq)+(m\sq-\ma\sq)
      [\Dl pp_{x}-\frac{1}{4}(m\sq-\ma\sq)]=0
    \eeq
  \item for $\Dl m\sq\neq0$:
    \begin{eqnarray}
      \lefteqn{\Dl E\sq p_{y}\sq+\Dl m\sq(p_{x}+\frac{\Dl p}{2}
               \frac{\Dl m\sq+m\sq-\ma\sq}{\Dl m\sq})\sq=}  \\
      & & \frac{\Dl E\sq}{\Dl m\sq}[\frac{1}{4}(\Dl m\sq+m\sq-\ma\sq)\sq
          -\Dl m\sq m\sq]  \nonumber
    \end{eqnarray}
\end{itemize}
In figures \ref{fig2} and \ref{fig3} two situations are described:
\begin{enumerate}
  \item $\Dl m\sq>0$: the intersection (when not empty) is bounded
    -- elliptic.
  \item $\Dl m\sq\leq0$: the intersection (when not empty) is unbounded
    however it is always bounded from the direction $-\Dl E\Dl\pv$
    (\ie in the direction of $\Dl\pv$ if $\Dl E$ is negative and in
    the opposite direction if $\Dl E$ is positive).
    \begin{itemize}
      \item For $\Dl m\sq=0$ this is a parabola.
      \item For $\Dl E\sq=0$ this is a straight line parallel to the
        $y$ axis.
      \item Otherwise it is a hyperbola (only one branch).
    \end{itemize}
\end{enumerate}
The width of the bands depends on $\rho$ and it may become infinite
for $\rho$ too large (\eg if $\Dl R$ contains the origin, $R$ will
obviously cover all of $\TF$). We choose $\rho$ small enough so that
the band will be bounded from the direction $-\Dl E\Dl\pv$.

In the following, $(\fA)\dg$ must be considered as well as $\fA$. Using
eq. (\ref{fAel}) one obtains:
\beq\label{fAdg}
  (\fA)\dg=f\dg\cdot A\dg
  \mbox{ where }f\dg(p):=\bbar{f(-p)},\;\forall p.
\eeq
Let $R\dg$ and $R_{m}\dg$ denote the regions in $\TF$ that correspond
to $(\fA)\dg$ as $R$ and $R_{m}$ to $\fA$.

Now it is possible to proceed with the proof. Let $p'\so\in\Tm$ for
some $m\in\M$.

\noindent{\bf Lemma 3:}
\begin{quote}\em
  There exist three different momenta $p\so,q\so,q'\so\in\Tm$ which
  are not in $\bbar{R}\cup\bbar{R\dg}$ \linebreak ($\bbar{R}$ denotes
  the closure of $R$ in $\TF$ and the same for $\bbar{R\dg}$) and
  satisfy \[  p\so+q\so=p'\so+q'\so  \]
\end{quote}
Proof:
\begin{quote}
  $R_{m}$ is a finite union of bands concentrated around
  $\Tm\cap(\TF-\Dl p)$.The analysis above showed that each band is
  bounded from the direction $-\Dl E\so\Dl\pv\so$ and thus so is
  $R_{m}$. To consider $R_{m}\dg$, all that is needed is to change
  $\Dl p\so\longrightarrow-\Dl p\so$ (according to eq. (\ref{fAdg}) ).
  This leaves $-\Dl E\so\Dl\pv\so$ unchanged so $R_{m}\dg$ is bounded
  from the same direction as $R_{m}$. Thus
  $\bbar{R_{m}}\cup\bbar{R_{m}\dg}$ is concentrated in a
  ``half-hypersurface'', and $p\so,q\so,q'\so$ can always be chosen
  in the other half. \qed
\end{quote}
With $p\so,q\so$ and $q'\so$ of lemma 3, we choose neighborhoods
$R_{p},R_{q}$ of $p\so$ and $q\so$ respectively in $\Tm$ such that
$R_{p},R_{q},R_{q'}:=(R_{p}+R_{q}-p'\so)\cap\Tm$ and
$\bbar{R_{m}}\cup\bbar{R_{m}\dg}$ are disjoint. It is easy to see that
such a choice is possible. Moreover, $R_{q'}$ thus defined is a
neighborhood of $q'\so$ in $\Tm$. (This is because the tangent
hyperplanes to $\Tm$ at $p\so$ and $q\so$ are not parallel so
$R_{p}+R_{q}-p'\so$ is a neighborhood of $q'\so$ in $\T$.)
 
Now let $\chi\in\Fone$ and denote $\f':=(\fA)\chi\;(\in\Foned)$.
 
\noindent{\bf Lemma 4:}
\begin{quote}\em
  If $\f,\psi,\psi'\in\Fone$ have support in $R_{p},R_{q}$ and
  $R_{q'}$ resp.\ then:
  \beq  <\f'\otimes\psi'|(S-I)(\f\otimes\psi))=0.  \eeq
\end{quote}
Proof:
\begin{quote}
  Since $R_{q}$ and $R_{q'}$ are disjoint, we have $(\psi'|\psi)=0$
  and thus
  \[  <\f'\otimes\psi'|I(\f\otimes\psi))=<\f'|\f)(\psi'|\psi)=0.  \]
  Since $R_{p},R_{q}$ and $R_{q'}$ don't intersect $R$ and $R\dg$,
  $\f,\psi$ and $\psi'$ are annihilated by $\fA$ and $(\fA)\dg$
  (according to lemma 2) so:
  \[  (\fA)(\chi\otimes\psi')=
      [(\fA)\chi]\otimes\psi'+\chi\otimes[(\fA)\psi']=
      \f'\otimes\psi'  \]
  and
  \[  (\fA)\dg(\f\otimes\psi)=0.  \]
  Combining this with the commutativity of $(\fA)\dg$ with $S$ we have:
  \begin{eqnarray*}
    \lefteqn{<\f'\otimes\psi'|S(\f\otimes\psi))=}  \\
    & & =<(\fA)(\chi\otimes\psi')|S(\f\otimes\psi))=  \\
    & & =(S^{*}(\chi\otimes\psi')|(\fA)\dg(\f\otimes\psi)>=0
  \end{eqnarray*}
  \qed
\end{quote}
{\bf Lemma 5:}  \begin{quote}  $\fA=0$.  \end{quote}
 
\noindent Proof:
\begin{quote}
  According to lemma 4 (and using eq. (\ref{SIT})
  \begin{eqnarray*}
    \lefteqn{\int d\mu(p')d\mu(q')d\mu(p)d\mu(q)\dl^{d}(p+q-p'-q')}  \\
    & & \f'(p')\bbar{\psi'(q')}<p',q'|T|p,q>\f(p)\psi(q)=0
  \end{eqnarray*}
  for each $\f,\psi,\psi'\in\Fone$ with support in
  $R_{p},R_{q},R_{q'}$ respectively. Suppose, by contradiction, that
  $p'\so$ is in supp$(\f')$. This implies that
  \[  <p'\so,p+q-p'\so|T|p,q>=0  \]
  whenever $p\in R_{p},q\in R_{q}$ and $p+q-p'\so\in R_{q'}$. This region
  corresponds to an open set in the $(s,t)$ plane, so the analyticity
  of $<T^{(2,2)}>$ (assumption 6 of the theorem) implies that it vanishes for
  any momenta in $\Tm$. In particular, the elastic scattering
  amplitudes vanish for particles with mass $m$, in contradiction to
  assumption 7 of the theorem.

  So $p'$ is not in the support of $\f'$. Since $p'$ is arbitrary,
  this means that $\f'=0$, \ie $\chi$ is annihilated by $\fA$ and
  since $\chi$ is arbitrary, this means that $\fA=0$. \qed
 
\end{quote}
Finally, summarizing this part, we have:
 
\noindent{\bf Proposition 1:}
\begin{quote}\em
  The momentum support of $<A>$ for any $A\in\A$ is restricted to the
  submanifold of $\TF\times\TF$ determined by the constraint $p'=p$.
\end{quote}
Proof:
\begin{quote}
  Lemma 5 showed that for each $0\neq\Dl p\in\T$ there is a
  neighborhood $\Dl R$ of $\Dl p$ such that for each $f\in\D(\T)$
  with support in $\Dl R$, $\fA=0$. Since
  \[  <p|\fA|p'>=f(p-p')<p|A|p'>,  \]
  this means that $A$ vanishes (as a distribution) at the region
  \[  \{(p,p')\in\TF\times\TF|p-p'\in\Dl R\}.  \]
  In particular, if $p'-p=\Dl p$ then $(p,p')$ is not in the support
  of $<A>$ and this is true for any $\Dl p\neq0$. \qed
\end{quote}

\subsection*{Part 2}

This part analyzes the structure of the elements of
\beq  \B:=\{B\in\A|B\mbox{ is self adjoint and }[P,B]=0\}  \eeq
\[  \mbox{where }[P,B]:=P'B-BP
    \mbox{ is an element of }L^{\times}(\Fone;\Foned)  \]
($P'$ is the dual of $P$ and is an element of $L(\F')$).
\beq  <p'|[P,B]|p>=(p'-p)<p'|B|p>  \eeq
thus for each self adjoint element $B$ of $\A$:
\[  B\in\B\;\Longleftrightarrow\;<p'|B|p>=\dl(p'-p)B(p)  \]
where $B(p)$ is a generalized function on the submanifold of
$\Om\times\Om$, characterized by the constraint $p'=p$
\cite[vol.\ 1, p.\ 209]{Gel'fand}, which is a matrix-valued generalized
function on $\TF$. As an operator (from $\Fone$ to $\Foned$) $B\in\B$
acts by multiplication:
\beq\label{Bop}  (B\f)(p)=B(p)\f(p)\;\forall\f\in\Fone  \eeq
(notice that this is a matrix multiplication, $B(p)$ being a Hermitian
matrix).
 
Most of the analysis will be performed on elements of
\beq  \Bi:=\{B\in\B|B\mbox{ is a smooth function of }p\in\TF\}.  \eeq
For each $p\in\TF$ and $B\in\Bi$, $B(p)$ is an operator in $\Hc(p)$
($N(p)$-dimensional matrix). According to eq. (\ref{Bop}), $B\in\Bi$
is a continuous operator in $\Fone$ and since it is self adjoint, it
is extended continuously by its dual $B'$ to $\Foned$. Its action on
``plane waves'' is:
\beq\label{Bpw}  B'|p>=|p>B(p)  \eeq
(recall from appendix \ref{Gelfand} that $|p>$ is a row vector)

\noindent{\bf Lemma 1:}
\begin{quote}\em
  $\Lc$ acts in $\B$ and $\Bi$ as a group of automorphisms.

  For each $B\in\B,\;(\Lm,a)\in\Pc$,
  \beq\label{UBUp}
    [U(\Lm,a)'BU(\Lm,a)](p)=\Lp(\Dl(\Lm,p'))\dg B(p')\Lp(\Dl(\Lm,p'))
  \eeq
  where $p':=\Lm p$.
\end{quote}
(Throughout this Section, ``$\dg$'' denotes the Hermitian
  conjugation of (matrix) maps between the spaces $\Hc(p^n)$.)
This follows directly from the expression for
\mbox{$<\psi|U(\Lm,a)'BU(\Lm,a)|\f>$}
($\f,\psi\in\Fone$), using also the $\Lc$-invariance of
$\mu$.

\subsubsection*{The Traceless Parts}

Let $B\in\Bi,\;p\in\TF$. The traceless part of $B(p)$ is
\beq\label{Bsp}  B^{*}(p)=B(p)-\frac{1}{N(p)}\one(p)\tr B(p)  \eeq
where (see appendix \ref{OneSpace} for review of notation)
\begin{itemize}
  \item $N(p)$ is the dimension of $\Hc(p)$;
  \item $\one(p)$ is the identity operator in $\Hc(p)$;
  \item the trace $\tr B(p)$ of $B(p)$ is:
    \[  \tr B(p):=
        \sum_{[\al,\lm]\in\Om(p)}B_{[\al,\lm][\al,\lm]}(p)  \]
\end{itemize}
(Notice that throughout this Section, ``$*$'' does not denote the
Hilbert-space-adjoint).
\noindent Let $\B^{*}$ denote the set of all traceless parts of
elements of $\Bi$:
\beq  \B^{*}:=\{B^{*}|B\in\Bi\}.  \eeq
{\bf Lemma 2:}
\begin{quote}\em
  $\Lc$ acts on $\B^{*}$ as a group of automorphisms: for each
  $\Lm\in\Lc,\;B\in\Bi$
  \beq  U(\Lm)'B^{*}U(\Lm)=[U(\Lm)'BU(\Lm)]^{*}.  \eeq
\end{quote}
Proof:
\begin{quote}
  Using eq.\ (\ref{UBUp}) we obtain:
  \begin{eqnarray*}
    \lefteqn{\tr[(U(\Lm)'BU(\Lm))(p)]=}  \\
    & & =\tr[\Lp(\Dl(\Lm,\Lm p))\dg B(\Lm p)\Lp(\Dl(\Lm,\Lm p))]
        =\tr[B(\Lm p)]
  \end{eqnarray*}
  therefore, using eq.\ (\ref{Bsp}), we have
  \begin{eqnarray}
    \lefteqn{[U(\Lm)'B^{*}U(\Lm)](p)=}  \\
    & & =\Lp(\Dl(\Lm,\Lm p))B^{*}(\Lm p)\Lp(\Dl(\Lm,\Lm p))
        =[U(\Lm)'BU(\Lm)]^{*}(p)  \nonumber
  \end{eqnarray}
  \qed
\end{quote}
In the next subsection it will be shown that this action is trivial.
To do this we show now that $\B^{*}$ is a Lie algebra of matrices. We say
that $p\sq\in\TF\sq$ is {\em a null pair} if there exists
$0\neq h\in\Hc(p\sq)$, such that \mbox{$<\!q\sq|T|p\sq\!>h$} vanishes as
a function of $q\sq\in\TF\sq$ (in physicists' terminology this implies
that ``plane waves'' with this combination of momenta, spins and particle
types do not scatter elastically). We now show that $B^{*}$ is determined
by its value for one non-null pair (and therefore $\B^{*}$ is, indeed, a
Lie algebra of matrices). This will be shown by stages, in the next four
lemmas.

As an element of $\A$, $B$ acts on two-particle states according to:
\beq
  B\f=(B\otimes I+I\otimes B)\f,\;\forall\f\in\Fone\times\Fone
\eeq
therefore
\beq  (B\f)(p,q)=B(p,q)\f(p,q)  \eeq
where
\beq  B(p,q):=B(p)\otimes\one(q)+\one(p)\otimes B(q).  \eeq
The trace of $B(p,q)$ is
\beq  \tr B(p,q)=N(q)\tr B(p)+N(p)\tr B(q)  \eeq
therefore the traceless part $B^{*}(p,q)$ of $B(p,q)$ is
\begin{eqnarray}
  B^{*}(p,q) & = & B(p,q)-\frac{1}{N(p)N(q)}
                   \one(p)\otimes\one(q)\tr B(p,q)  \nonumber\\
             & = & B^{*}(p)\otimes\one(q)+\one(p)\otimes B^{*}(q).
\end{eqnarray}
which means that
\beq\label{Bspq}
  B^{*}(p,q)=0\;\Longleftrightarrow\;
  B^{*}(p)=0\mbox{ and }B^{*}(q)=0.
\eeq
{\bf Lemma 3:}
\begin{quote}
  If for $p,q\in\TF$, $(p,q)$ is a non-null pair and $B^{*}(p,q)=0$
  then for each $\Lm\in\Lc(p+q),\;B^{*}(\Lm p,\Lm q)=0$
\end{quote}
Proof:
\begin{quote}
  Let $\Lm\in\Lc(p+q)$ (a ``rotation'' in the ``center of mass'' of
  $(p,q)$). We denote $p':=\Lm p,\;q':=\Lm q$. Since $B^{*}(p,q)=0$,
  $B(p,q)$ is a scalar matrix so it commutes with
  $\Lp(\Dl(\Lm,p'))\otimes L^{(q)}(\Dl(\Lm,q'))$. Thus eq. (\ref{UBUp})
  reduces in this case to
  \beq\label{BpqLm}  B(p',q')=[U(\Lm\inv)BU(\Lm)](p,q).  \eeq
  As an element of a Lie group, $\Lm$ belongs to some one-parameter
  subgroup $\Lm(\th)$ of $\Lc(p+q)$ (analytic in $\th$) which
  is generated in $\Fone$ by a continuous operator $J$ (see appendix
  \ref{Gen}):
  \beq  U(\Lm(\th))\f=e^{i\th J}\f,\;\forall\f\in\Fone.  \eeq
  A Taylor expansion in $\th$ gives:
  \beq\label{Taylor}
    U(\Lm(\th)\inv)BU(\Lm(\th))=e^{-i\th J}Be^{i\th J}
    =\sum_{n=0}^{\infty}\frac{\th\n}{n!}\FJ\n(B)
  \eeq
  (the left hand side is a holomorphic (operator valued)
  function of $\th$, thus the right hand side converges absolutely
  for all $\th$) where $\FJ$ is an operator on $L(\Fone)$ defined by
  \beq  \FJ(C):=i[J,C],\;\forall C\in L(\Fone)  \eeq
  In the following we prove that if $C\in\Bi$ satisfies $C^{*}(p,q)=0$
  then so does $\FJ(C)$. This implies, by induction, that for each
  integer $n$, $[\FJ\n(B)](p,q)$ is scalar. Combining eqs.
  (\ref{BpqLm}) and (\ref{Taylor}) we conclude that $B(p',q')$ is
  scalar, which means that $B^{*}(p',q')=0$, as claimed. So it is left
  to show that $[J,B]^{*}(p,q)=0$.
 
  $B(\Lm(\th)p)\equiv(e^{-i\th J}Be^{i\th J})(p)$ is a smooth
  function of $\th$ and $p$ so
  \[  i[J,B](p)=\der{\th}(e^{-i\th J}Be^{i\th J})|_{\th=0}  \]
  is also smooth in $p$. Since $B$ and $J$ are self adjoint, so is
  $i[J,B]$ and according to property 4 of $\A$, $i[J,B]$ is an element
  of $\A$. So we can conclude that $i[J,B]\in\Bi$.
 
  Suppose, that $B^{*}(p',q')\neq0$. $B(p',q')$ is
  diagonalizable (since it is Hermitian) and since it is not scalar,
  it  has at least two different eigenvalues, so at least one of them
  $b'$ is different from the (unique) eigenvalue $b$ of $B(p,q)$.
  Let $h\in\Hc(p,q)$ be an eigenvector of $B(p',q')$ belonging to $b'$.
  $B$ commutes with $S$ (as a self adjoint element of $\A$) so we have
  (using eq. (\ref{Bpw}))
  \begin{eqnarray*}
    0 & = & h\dg<p',q'|[B,S]|p,q>h=  \\
      & = & [B(p',q')h]\dg<p',q'|S|p,q>h
            -h\dg<p',q'|S|p,q>[B(p,q)h]=  \\
      & = & (b'-b)h\dg<p',q'|S|p,q>h.
  \end{eqnarray*}
  For $\Lm\neq1$ this becomes
  \beq\label{hTh}  h\dg<p',q'|T|p,q>h=0  \eeq
  (Here we use the fact that $\Lm$ is in $\Lc(p+q)$ which means that 
  $p'+q'=p+q$) so the assumption that $B^{*}(p',q')\neq0$ leads to eq. 
  (\ref{hTh}).
  We are going to show that eq. (\ref{hTh}) cannot be satisfied for
  $\Lm=1$. This implies, since the left-hand side of (\ref{hTh}) is
  continuous in $\th$, that (\ref{hTh})  is not satisfied also in some
  neighborhood of the identity so in this neighborhood $B^{*}(p',q')=0$.
  \begin{eqnarray*}
    B(p',q') & = & (e^{-i\th J}Be^{i\th J})(p,q)= \\
             & = & B(p,q)+i\th[J,B](p,q)+{\cal O}(\th\sq)
  \end{eqnarray*}
  Since $B^{*}(p,q)=0$ and for $\th$ sufficiently small
  $B^{*}(p',q')=0$, we have \linebreak $[J,B]^{*}(p,q)=0$.

  Finally we must show that eq. (\ref{hTh}) cannot be satisfied for
  $\Lm=1$, \ie
  \beq h\dg<\!p\sq|T|p\sq\!>h\neq0 \eeq
  for any $0\neq h\in\Hc(p^2)$, where $p^2=(p,q)$.
  The expression
  \beq\label{int}  h\dg<\!p\sq|T|q\m\!>\dg<\!q\m|T|p\sq\!>h  \eeq
  is non negative (being the norm of $<\!q\m|T|p\sq\!>h$). Since $p\sq$
  is non null, there exists $q_{0}\sq\in\TF\sq$ for which
  $<\!q_{0}\sq|T|p\sq\!>h\neq0$, and since $<\!T^{(2,2)}\!>$ is
  analytic, the expression (\ref{int}) is positive in some
  neighborhood of $q_{0}\sq$. Using the optical theorem (see appendix
  \ref{scatter}) we, therefore, obtain
  \begin{eqnarray}
    \nonumber  0 & < & \int_{\TF^{m_{0}}}d\mu\sq(q\sq)(2\pi)^{d}\dl^{d}
                        (q_{1}+q_{2}-p_{1}-p_{2})
                        h\dg<\!p\sq|T|q\sq\!>\dg<\!q\sq|T|p\sq\!>h  \\
    \label{pos}  & \leq & \frac{1}{i}h\dg(<\!p\sq|T|p\sq\!>
                           -<\!p\sq|T|p\sq\!>\dg)h  \\
    \nonumber    & = & 2\mbox{Im}(h\dg<\!p\sq|T|p\sq\!>h).
  \end{eqnarray}
  \qed
\end{quote}
For $p_{1},p_{2}\in\Tm$, let $S(p_{1},p_{2})$ be the set of all the
space-like parts of elements of $\Tm$ that can be transformed to
$p_{1}$ by an element of $\Lc(p_{1}+p_{2})$. If $p_{1}+p_{2}=0$ then
$\Lc(p_{1}+p_{2})$ is the rotation group so $S(p_{1},p_{2})$ is a
sphere around the origin containing $\pv_{1}$ and $\pv_{2}$. In
general, $S(p_{1},p_{2})$ is the transformation of such a sphere by
the $\Lc$-transformation $\Lm_{p_{1}+p_{2}}$, and it can be shown that
it is a ($r-1$ dimensional) ellipsoid, symmetric around
$\pv_{1}+\pv_{2}$ and longer in this direction then in the other ones.
 
\noindent{\bf Lemma 4:}
\begin{quote}\em
  If for $p_{1},p_{2}\in\Tm$, $(p_{1},p_{2})$ is non-null and
  $B^{*}(p_{1},p_{2})=0$ then for each \linebreak
  $q_{1}\in S(p_{1},p_{2})$, $B^{*}(q_{1})=0$.
\end{quote}
Proof:
\begin{quote}
  By definition, $q_{1}\in S(p_{1},p_{2})$ iff $q_{1}\in\Tm$ and there
  exists $q_{2}\in\Tm$ for which \mbox{$q_{1}+q_{2}=p_{1}+p_{2}$}
  ($q_{2}$ is the momentum transformed to $p_{2}$ by the element of
  \linebreak $\Lc(p_{1}+p_{2})$ which transforms $q_{1}$ to $p_{1}$).
  Since
  $(p_{1},p_{2})$ is non-null and \linebreak $B^{*}(p_{1},p_{2})=0$,
  lemma 3 implies that $B^{*}(q_{1},q_{2})=0$ so according to
  relation (\ref{Bspq}), \mbox{$B^{*}(q_{1})=0$.} \qed
\end{quote}
{\bf Lemma 5:}
\begin{quote}\em
  If, for $r>0$, each $p\in\Tm$ with $|\pv|=r$ satisfies $B^{*}(p)=0$
  then there exists $r'>r$ such that for each $p'\in\Tm$ which
  satisfies $|\pv\,'|\leq r'$, $B^{*}(p')=0$.
\end{quote}
Proof:
\begin{quote}
  We construct $p_{1},p_{2}\in\Tm$ with $|\pv_{1}|=|\pv_{2}|=r$ for
  which $\psm=(m,0,\ldots,0)$ is in $S(p_{1},p_{2})$. This requirement
  is equivalent to
  \beq  p_{1}+p_{2}-\psm\in\Tm  \eeq
  which gives a condition on the angle $\th$ between $\pv_{1}$ and
  $\pv_{2}$:
  \beq
    2\sqrt{m\sq+r\sq}=m+\sqrt{m\sq+4r\sq\cos\sq\frac{\th}{2}}.
  \eeq
  This equation has a (unique positive) solution for $\th$ which is an
  increasing function of $\frac{r}{m}$. Thus the required momenta
  exist and they satisfy \linebreak $\pv_{1}\neq\pv_{2}\neq-\pv_{1}$
  $(0\neq\th\neq\pi)$.
 
  Now define
  \beq  r':=\max\{|\pv\,'|\;|\;\pv\,'\in S(p_{1},p_{2})\}  \eeq
  (the maximum exists since $\Lc(p_{1}+p_{2})$ is compact and acts
  continuously in $\Tm$). Considering the shape of $S(p_{1},p_{2})$,
  the fact that $\pv_{1}\neq\pv_{2}$ implies that $S(p_{1},p_{2})$
  is not contained entirely inside this sphere which means that
  $r'>r$. $S(p_{1},p_{2})$ is connected (this comes from the
  connectedness of $\Lc(p_{1}+p_{2})$) thus it contains momenta with
  any magnitude in the range $[0,r']$.
 
  Let $p'\in\Tm$ for which $|\pv\,'|\leq r'$. One can always perform a
  rotation around the origin, transforming $p_{1}$ and $p_{2}$ to put
  $\pv\,'$ in $S(p_{1},p_{2})$. This rotation does not change the
  magnitude of $\pv_{1}$ and $\pv_{2}$ so we still have
  $B^{*}(p_{1})=B^{*}(p_{2})=0$ which means (by eq. (\ref{Bspq})) that
  $B^{*}(p_{1},p_{2})=0$. If $(p_{1},p_{2})$ is non-null then lemma 4
  implies that $B^{*}(p')=0$. If $(p_{1},p_{2})$ is null, a slight
  deformation of $p_{2}$ preserving $|\pv_{2}|=r$ can be made to
  change $s=(p_{1}+p_{2})\sq$ and since null pairs exist only for
  isolated values of $(s,t)$ (according to assumptions 6 and 7),
  we will get a non-null pair. This implies that
  $p'$ is on a boundary of a region in which $B^{*}$ vanishes so, by
  the continuity of $B^{*}$, we obtain $B^{*}(p')=0$. \qed
\end{quote}
{\bf Lemma 6:}
\begin{quote}\em
  If $p,q\in\Tm$ are different and such that $(p,q)$ is non-null and
  \mbox{$B^{*}(p,q)=0$} then $B^{*}$ vanishes on $\Tm$.
\end{quote}
Proof:
\begin{quote}
  First assume $\pv+\vec{q}=0$ and define
  \beq
    r:=\sup\{r'|B^{*}(p')=0,
       \mbox{ for all }p'\in\Tm\mbox{ with }|\pv\,'|\leq r'\}.
  \eeq
  Lemma 4 implies that for each $p'\in\Tm$ with $|\pv\,'|=|\pv|$,
  $B^{*}(p')=0$, so, by lemma 5, $r\geq|\pv|>0$.
 
  Assume, by contradiction, that $r$ is finite. By definition, for each
  \mbox{$p'\in\Tm$} with $|\pv\,'|<r$, $B^{*}(p')=0$ so if $p'\in\Tm$
  satisfies $|\pv\,'|=r$, it is on the boundary of a region in which
  $B^{*}$ vanishes, thus, by continuity, $B^{*}(p')=0$. Looking back at
  lemma 5 one observes that it contradicts the maximality of $r$.
  Therefore $r=\infty$ which means that $B^{*}$ vanishes on $\Tm$.
 
  Returning to the general case (where $\pv+\vec{q}\neq0$), define:
  \[  p':=\Lm_{p+q}\inv p,\;q':=\Lm_{p+q}\inv q,\;
      B':=U(\Lm_{p+q}\inv)BU(\Lm_{p+q})  \]
  (this is a transformation to the ``rest frame'' of $p+q$). Eq.
  (\ref{Bspq}) implies that $B(p)$ and $B(q)$ are scalars thus commute
  with $\Lp$. Therefore, \mbox{eq. (\ref{UBUp})} gives
  \[ B'(p')=\Lp(\Dl(\Lm_{p+q},p))\dg B(p)\Lp(\Dl(\Lm_{p+q},p))=B(p) \]
  and in the same way $B'(q')=B(q)$, thus
  \[  B'^{*}(p',q')=B^{*}(p,q)=0.  \]
  The pair $(p',q')$ is non-null since it was obtained from a non-null
  pair by an
  $\Lc$-transformation which doesn't change the S-matrix; therefore
  $B'$ satisfies the assumptions of the lemma and $\pv\,'+\vec{q}\,'=0$,
  so the first part implies that $B'^{*}=0$. Since
  $B^{*}=U(\Lm_{p+q})B'^{*}U(\Lm_{p+q}\inv)$ (according to lemma 2)
  this means that also $B^{*}=0$. \qed
\end{quote}
>From the property 4 of $\A$ and the definitions of $\B,\Bi$, and
$\B^{*}$, it follows that $\B^{*}$ is a real Lie algebra. If
$B_{1}$ and $B_{2}$ are in $\Bi$ then so is $i[B_{1},B_{2}]$ and
\beq  i[B_{1}^{*},B_{2}^{*}]=i[B_{1},B_{2}]^{*}\in\B^{*}.  \eeq
Denote by $\Bsm$ the set of restrictions of elements of $\B^{*}$ to
one orbit $\Tm$ ($m\in\M$). This is a real Lie algebra of smooth
functions from $\Tm$ to $\Hc(m)$. For $B^{*}\in\Bsm$, \mbox{$p\in\Tm$},
$B^{*}(p)$ is a traceless Hermitian $N(m)\times N(m)$ dimensional
matrix which is an element of the Lie algebra $su(N(m))$ of the group
$SU(N(m)$. Therefore, for $p,q\in\Tm$, $B^{*}(p,q)$ can be identified
as an element of $su(N(m))\oplus su(N(m))$ which is a finite
dimensional compact Lie algebra. The mapping
\beq  B^{*}\mapsto B^{*}(p,q)  \eeq
is a homomorphism for the Lie algebra structure and lemma 6 implies
that if $(p,q)$ is non-null, this homomorphism is an isomorphism
(injective) which implies (since non-null pairs exist) that $\Bsm$ can
be identified as a subalgebra
of a compact Lie algebra (of matrices). As such, its structure is
\beq  \Bsm=\Bs\oplus\Bcc\mbox{ (a direct sum of ideals)}  \eeq
where
\begin{description}
  \item $\Bs$ is a semisimple compact Lie algebra;
  \item $\Bcc$ is the center of $\Bsm$, an abelian Lie algebra.
\end{description}

\subsubsection*{The Action of $\Lc$ on the Traceless Parts}

According to lemmas 1 and 2, $\Lc$ acts on $\Bsm$ as a group of
automorphisms and this action preserves the commutator:
\beq\label{UBUcom}
  [U\inv B_{1}U,U\inv B_{2}U]=U\inv[B_{1},B_{2}]U,\;
  \forall B_{1},B_{2}\in\Bsm.
\eeq
 
We use this to prove:
 
\noindent{\bf Lemma 7:}
\begin{quote}\em
  $\Bs$ and $\Bcc$ are $\Lc$-invariant
\end{quote}
Proof:
\begin{quote}
  For sets $S_{1},S_{2}\subset\Bsm$, denote $[S_{1},S_{2}]$ the real
  vector space spanned by
  \beq  \{[B_{1},B_{2}]|B_{1}\in S_{1},B_{2}\in S_{2}\}.  \eeq
  $U\Bsm U\inv=\Bsm$ and $[\Bcc,\Bsm]=0$ so
  \beq  [U\inv\Bcc U,\Bsm]=U\inv[\Bcc,U\Bsm U\inv]U=0  \eeq
  thus  $U\inv\Bcc U\subset\Bcc$ \ie $\Bcc$ is $\Lc$-invariant.

  $[\Bcc,\Bsm]=0$ implies that $[\Bsm,\Bsm]=[\Bs,\Bs]$ and the
  semisimplicity of $\Bs$ implies that $[\Bs,\Bs]=\Bs$; therefore
  \begin{eqnarray*}
    U\inv\Bs U & = & U\inv[\Bsm,\Bsm]U=[U\inv\Bsm U,U\inv\Bsm U]=  \\
               & = & [\Bsm,\Bsm]=\Bs,
  \end{eqnarray*}
  which means that $\Bs$ is $\Lc$-invariant. \qed
\end{quote}
Lemma 7 means that $\Lc$ acts as a group of automorphisms in each of
the ideals $\Bs$ and $\Bcc$. The representation of $\Lc$ (as a group of
automorphisms) is a homomorphism so its kernel is a normal subgroup of
$\Lc$. $\Lc\so$ is a simple group which means that it doesn't have
non-trivial normal subgroups (that is other than $\Lc\so$ itself and
$\{e\}$, where $e$ is the identity in $\Lc\so$).
$\Lc\so=\Lc/{\bf Z}_{2}$ (since $r\geq3$; see appendix \ref{Poinkare}),
Therefore ${\bf Z}_{2}$ is the only non trivial normal subgroup
of $\Lc$. To show that $\Lc$ is represented
trivially (the kernel is $\Lc$) it is enough to show that the kernel
is not contained in ${\bf Z}_{2}$. This will be
done in the next two lemmas.

\noindent{\bf Lemma 8:}
\begin{quote}\em  $\Lc$ acts trivially in $\Bs$.  \end{quote}
Proof:
\begin{quote}
  The connected part of the group of automorphisms of a compact
  semisimple Lie algebra is known to be the corresponding compact
  semisimple group, so the representation of $\Lc$ in $\Bs$ is a
  homomorphism from $\Lc$ to a compact group. Since $\Lc$ and
  $\Lc\so=\Lc/{\bf Z}_{2}$ are not compact and therefore cannot be
  contained in a compact group, the kernel must be all of $\Lc$. \qed
\end{quote}
{\bf Lemma 9:}
\begin{quote}\em  $\Lc$ acts trivially in $\Bcc$.  \end{quote}
Proof:
\begin{quote}
  We choose $p,q\in\Tm$ such that $(p,q)$ is non-null and
  $\pv+\vec{q}=0$ (so that \mbox{$\Lc(p+q)={\cal O}_{0}(r)$}, the
  connected part of ${\cal O}(r)$). Since $r\geq3$, $\Lc(p+q)$ contains
  a one-parameter subgroup of rotations $R(\th)$ around $\pv$. $R(\th)$
  doesn't change $p$ and $q$ therefore it is represented in $\Bsm$ by
  $\Lp\otimes L^{(q)}$ acting on the matrices $B^{*}(p,q)$. $R(\th)$
  is abelian thus its (finite dimensional) irreducible representations
  in a complex Hilbert space are one-dimensional. Let $B$ be an
  element of the complex extension of $\Bcc$ which transforms
  irreducibly under $R(\th)$. In the following we will show that if $B$
  doesn't transform trivially then $B$ cannot commute with its
  adjoint, in contradiction to the commutativity of $\Bcc$. To show
  this, we need an explicit representation for the matrix $B(p,q)$.
  Let $J$ be the generator of $(L^{(p)}\otimes L^{(q)})(R(\th))$.
  This is a Hermitian matrix (since $\Lp$ is unitary) thus
  diagonalizable, so we assume that the diagonalization has been
  performed. In this basis, $B(p,q)$, abbreviated in the following by
  $B$, can be seen as a block matrix $B=\{B_{ij}\}$, where each block
  $B_{ij}$ connects the eigenspaces of the eigenvalues $i$ and $j$ of
  $J$. The action of $R(\th)$ on $B$ is
  \beq\label{RthB}
    B\stackrel{R(\th)}{\mapsto}e^{-i\th J}Be^{i\th J}=e^{-i\th k}B
  \eeq
  where k is some real number. By optionally switching $B$ with
  $B\dg$ one can always have $k\geq0$ so we assume that it is so.
  Differentiating \mbox{eq. (\ref{RthB})}, one obtains:
  \[  [J,B]=kB.  \]
  This means that B is a ``ladder operator'' for $J$: if
  $h\in\Hc(p,q)$ satisfies \mbox{$Jh=jh$} then $J(Bh)=(j+k)(Bh)$, and
  this implies that $B_{ij}$ can be non-zero only if \mbox{$i-j=k$}.
  $(B\dg)_{ij}=(B_{ji})\dg$ so we have:
  \begin{eqnarray}
    (BB\dg)_{ij}  & = & \sum_{l}B_{il}(B_{jl})\dg
                        =B_{i,i-k}(B_{j,j-k})\dg  \nonumber\\
                  & = & \dl_{ij}B_{i,i-k}(B_{j,j-k})\dg  \label{BBd}\\
    \mbox{and in the same way}  \nonumber\\
    (B\dg B)_{ij} & = & \dl_{ij}(B_{i+k,i})\dg B_{i+k,i} \label{BdB}.
  \end{eqnarray}
  Let $l$ be the maximal eigenvalue of $J$ for which $B_{l,l-k}\neq0$
  ($\Hc(p,q)$ is finite-dimensional and $B\neq0$ so the maximum exists).
  This implies that \linebreak $B_{l,l-k}(B_{l,l-k})\dg\neq0$ (since
  each element is a sum of squares). Now the commutativity of $\Bcc$
  implies that $BB\dg=B\dg B$ therefore eqs. (\ref{BBd}) and (\ref{BdB})
  give
  \beq  (B_{l+k,l})\dg B_{l+k,l}=B_{l,l-k}(B_{l,l-k})\dg\neq0.  \eeq
  This means that $B_{l+k,l}\neq0$ and the maximality of $l$ implies
  that $k=0$ which means that $B$ transforms trivially. Since $B$ is
  arbitrary, this implies that $J$ acts trivially which means that
  the kernel of the representation contains a one parameter group,
  which cannot be contained in ${\bf Z}_{2}$. Therefore the
  kernel is all of $\Lc$. \qed
\end{quote}

\subsubsection*{The General Form of the Elements of $\B$}

The triviality of the action of $\Lc$ in $B^{*}$ (lemmas 7,8 and 9)
implies that:
 
\noindent{\bf Lemma 10:}
\begin{quote}\em
  For each $B\in\Bi,\;p\in\Tm$, the traceless part has the form
  \beq\label{Bscom}
    B^{*}_{[\al\lm][\al'\lm']}=B^{*}_{\al\al'}\dl_{\lm\lm'}
  \eeq
  \ie $B^{*}$ is independent of $p$ and it commutes with $\Lp$.
\end{quote}
Proof:
\begin{quote}
  Lemmas 7,8 and 9 imply that for each $\Lm\in\Lc$
  \beq  U(\Lm\inv)B^{*}U(\Lm)=B^{*}  \eeq
  which means (by eq. (\ref{UBUp})) that for each $p\in\Tm$
  \beq\label{Bspp}
    B^{*}(p)=\Lp(\Dl(\Lm,p'))\dg B^{*}(p')\Lp(\Dl(\Lm,p')) \hsc
    (p'=\Lm p).
  \eeq
  Recall (from appendix \ref{RepAl}) that
  \beq\label{DLp} \Dl(\Lm,p)=\Lmp\inv\Lm\Lm_{\Lm\inv p} \eeq
  where
  \[ \Lmp \psm=p \hsc \Lm_{\psm}=1. \]
  From this it follows that $\Dl(\Lmp,p)=1$. Using this in eq.
  (\ref{Bspp}) gives
  \beq
    B^{*}(\psm)=\Lp(\Dl(\Lmp,p))\dg B^{*}(p)\Lp(\Dl(\Lmp,p))
               =B^{*}(p)
  \eeq
  therefore $B^{*}$ is independent of $p$. Eq. (\ref{DLp}) also implies
  that
  \beq  \Dl(\Lm,\psm)=\Lm,\;\forall\Lm\in\Lc(\psm).  \eeq
  Using this in eq. (\ref{Bspp}) gives (since $B^{*}(p')=B^{*}(p)$)
  \beq  [B^{*},\Lp(\Dl)]=0,\;\forall\Dl\in\Lc(\psm)  \eeq
  and eq. (\ref{Bscom}) follows from this by Schur's lemma. \qed
\end{quote}
{\bf Lemma 11:}
\begin{quote}\em
  For each $B\in\Bi$, if $p,q,p',q'\in\Tm$ satisfy $p+q=p'+q'$ then
  \em\beq  \tr B(p)+\tr B(q)=\tr B(p')+\tr B(q')  \eeq\em
  which means that {\em$\tr B(p)$} is a linear (real) function of $p$:
  \em\beq\label{trBp}  \tr B(p)=a_{\mu}p^{\mu}+b  \eeq
\end{quote}
Proof:
\begin{quote}
  Lemma 10 implies that
  \beq\label{Bpq10}
    B(p,q)=B^{*}(p,q)+\frac{1}{N(m)\sq}\one(p,q)\tr B(p,q)
  \eeq
  where $B^{*}$ is some constant Hermitian matrix. Following a similar
  procedure as in the proof of lemma 3, we take some eigenvector
  $h\in\Hc(p,q)$ of $B^{*}$ belonging to some eigenvalue $b$.
  Eq. (\ref{Bpq10}) implies that for each $p,q\in\Tm$, $h$ is an
  eigenvector of $B$, with the eigenvalue
  \beq
    b+\frac{1}{N(m)\sq}\tr B(p,q)=b+\frac{1}{N(m)}(\tr B(p)+\tr B(q)).
  \eeq
  Therefore we obtain
  \begin{eqnarray*}
    0 & = & h\dg<p',q'|[B,S]|p,q>h=  \\
      & = & \frac{1}{N(m)}[\tr B(p')+\tr B(q')-\tr B(p)-\tr B(q)]
            h\dg<p',q'|S|p,q>h
  \end{eqnarray*}
  and as in lemma 3, this implies that for a non-null pair,
  $\tr B(p)+\tr B(q)$ is locally $\Lc(p+q)$-invariant. Since $\Lc(p+q)$
  is connected, this means global invariance and since each null pair
  is a limit of non-null pairs and $B(p)$ is continuous in $p$, this
  holds also for null pairs. \qed
\end{quote}
To extend these results from $\Bi$ to $\B$, we define, for each
$B\in\B,\;f\in C_{c}^{\infty}(\Lc)$
\beq  B^{f}:=\int_{\Lc}d\Lm f(\Lm)U(\Lm)'BU(\Lm)  \eeq
where $d\Lm$ is the Haar measure of $\Lc$.

\noindent{\bf Lemma 12:}
\begin{quote}\em
  For each $B\in\B,\;f\in C_{c}^{\infty}(\Lc)$, $B^{f}$ is an element
  of $\Bi$.
\end{quote}
Proof:
\begin{quote}
  Let $\f,\psi\in\Fone$.
  \begin{eqnarray*}
    \lefteqn{(\psi|B^{f}\f>=}  \\
    & = & \int_{\Lc}d\Lm f(\Lm)\int_{\TF}d\mu(p)\psi(\Lm\inv p)\dg
          \Lp(\Dl(\Lm,p))\dg B(p)\Lp(\Dl(\Lm,p))\f(\Lm\inv p)=  \\
    & = & \sum_{m\in\M}\int_{\Tm}d\mu(p)
          \int_{\Lc}d(\Lmp\Lm)f(\Lmp\Lm)  \\
    &   & \;\;\psi(\Lm\inv\psm)\dg\Lp(\Dl(\Lmp\Lm,p))\dg B(p)
          \Lp(\Dl(\Lmp\Lm,p))\f(\Lm\inv\psm)
  \end{eqnarray*}
  where $\psm=\Lmp\inv p$ (the last equality is due to Fubini's
  theorem for distributions and the fact that multiplication by $\Lmp$
  is a bijection in $\Lc$). From eq. (\ref{BDLp}) we obtain
  \beq
    \Dl(\Lmp\Lm,p)=\Lm\Lm_{\Lm\inv\psm}\mbox{ (independent of $p$)}
  \eeq
  therefore (using the $\Lc$-invariance of $d\Lm$ and recalling that
  $\Lp$ is determined by $m$)
  \begin{eqnarray*}
    \lefteqn{(\psi|B^{f}\f>=}  \\
    & = & \sum_{m\in\M}\int_{\Lc}d\Lm
          \psi(\Lm\inv\psm)\dg\Lp(\Lm\Lm_{\Lm\inv\psm}\dg)  \\
    &   & \;[\int_{\Tm}d\mu(p)B(p)f(\Lmp\Lm)]
          \Lp(\Lm\Lm_{\Lm\inv\psm})\f(\Lm\inv\psm)  \\
    & = & \sum_{m\in\M}\int_{\Lc}d\Lm
          \psi(\Lm\psm)\dg\Lp(\Lm\inv\Lm_{\Lm\psm}\dg)  \\
    &   & \;[\int_{\Tm}d\mu(p)B(p)f(\Lmp\Lm\inv)]
          \Lp(\Lm\inv\Lm_{\Lm\psm})\f(\Lm\psm)
  \end{eqnarray*}
  Each $\Lm\in\Lc$ can be decomposed uniquely to $\Lm=\Lmp\Dl$ where
  $p=\Lm\psm$ and $\Dl=\Lmp\inv\Lm\in\Lc(\psm)$ therefore
  \[  \int_{\Lc}d\Lm\ldots=
      \int_{\Tm}d\mu(p)\int_{\Lmp\Lc(\psm)}d\Lm\ldots  \]
  and this gives:
  \beq  (\psi|B^{f}\f>=\int_{\TF}d\mu(p)\psi(p)\dg B^{f}(p)\f(p)  \eeq
  where
  \begin{eqnarray*}
    \lefteqn{B^{f}(p)=}  \\
    & & \int_{\Lmp\Lc(\psm)}d\Lm\Lp(\Lm\inv\Lmp)\dg
        [\int_{\Tm}d\mu(p')B(p')f(\Lm_{p'}\Lm\inv)]\Lp(\Lm\inv\Lmp)
  \end{eqnarray*}
  Now the smoothness of $B^{f}(p)$ follows from the analyticity of
  $\Lc$ and the smoothness of $f,\Lmp$ and $\Lp$. \qed
\end{quote}
Now it is possible to conclude with

\noindent{\bf Proposition 2:}
\begin{quote}\em
  On each orbit $\Tm$, $B\in\B$ has the form
  \beq\label{Bpr}  B(p)=B+Ia_{\mu}p^{\mu}  \eeq
  where $B$ is a Hermitian matrix of the form
  \[  B_{[\al\lm][\al'\lm']}=B_{\al\al'}\dl_{\lm\lm'}  \]
  and $\{a_{\mu}\}$ is a real vector.
\end{quote}
Proof:
\begin{quote}
  For $B\in\Bi$ it is a direct result of lemmas 10 and 11. Lemma 12
  implies that for each $f\in C_{c}^{\infty}(\Lc)$, $B^{f}$ is in $\Bi$
  so it has the form (\ref{Bpr}), therefore so does $B$. \qed
\end{quote}

\subsection*{Part 3}

In this part, the results of the preceding parts (propositions 1 and 2)
are combined to prove the statement of the theorem.

Let $A\in\A$. Proposition 1 states that the support of $<A>$ is
restricted to the submanifold of $\TF\times\TF$ defined by the
constraint $p'-p=0$. This implies \mbox{\cite[vol. 1]{Gel'fand}} that
$<A>$ is a polynomial in the derivatives of $\dl(p'-p)$. In other words,
$<A>$ is a differential operator (of finite order $N$):
\beq\label{Apol}
  <p'|A|p>=\dl(p'-p)\sum_{n=0}^{N}A_{\vec{\mu}}\ns(p)
           \frac{1}{i^{n}}\derp{p^{\mu_{1}}}\cdots\derp{p^{\mu_{n}}}
\eeq
where the coefficients $A_{\vec{\mu}}\ns(p)$ are generalized (matrix
valued) functions on $\TF$ and implicit summation over
$\vec{\mu}=(\mu_1,\ldots,\mu_n)$ is assumed.
Moreover, $\TF$ consists of {\em isolated}
orbits (since the one-particle mass spectrum is finite), therefore the
derivatives in $A$ are tangent to the orbits (which means that on each
orbit $\Tm$, $A$ is a polynomial in
$\nabla_{\mu}=\derp{p^{\mu}}-\frac{p^{\mu}p^{\nu}}{m\sq}\derp{p^{\nu}}$)
and this implies that
\beq\label{AcomP}  [A,P_{\mu}P^{\mu}]=0. \eeq
Combining eqs. (\ref{Apol}) and (\ref{AcomP}) with proposition 2, we
obtain:
 
\noindent{\bf Lemma 1:}
\begin{quote}\em
  For each self adjoint element $A$ of $\A$, $A_{\vec{\mu}}\Ns$
  belongs to $\B$. As such, it has the form
  \beq\label{ANp}
    A_{\vec{\mu}}\Ns(p)=Ia_{\nu\vec{\mu}}p^{\nu}+B_{\vec{\mu}},\;
    B_{\vec{\mu}[\al\lm][\al'\lm']}=B_{\vec{\mu}\al\al'}\dl_{\lm\lm'}
  \eeq
  and if $N\geq1$ then
  \beq\label{ind}
    B_{\vec{\mu}}=0,\;a_{\nu\mu_{1}\mu_{2}\cdots\mu_{N}}
                      =-a_{\mu_{1}\nu\mu_{2}\cdots\mu_{N}}.
  \eeq
\end{quote}
Proof:
\begin{quote}
  From $[A_{\vec{\mu}}\ns,P_{\nu}]=0$ and
  $[\derp{p_{\mu}},P_{\nu}]=\dl_{\nu}^{\mu}$ it follows that
  \[  i[i[\ldots i[A,P_{\mu_{N}}]\ldots,P_{\mu_{2}}],P_{\mu_{1}}]
      =A_{\vec{\mu}}\Ns.  \]
  Property 4 of $\A$ implies that $A_{\vec{\mu}}\ns$ is in $\A$. $A$
  and $P_{\mu}$ are Hermitian thus so is $A_{\vec{\mu}}\Ns$. Finally
  $[A_{\vec{\mu}}\Ns,P]=0$ therefore $A_{\vec{\mu}}\Ns$ is in
  $\B$ and \mbox{proposition 2} gives its general form (\ref{ANp}).
 
  For $N\geq0$, property 4 of $\A$ implies that
  \[  [[\ldots[A,P_{\mu_{N}}]\ldots,P_{\mu_{3}}],P_{\mu_{2}}]\in\A  \]
  therefore, from eq. (\ref{AcomP}) it follows that
  \begin{eqnarray}\label{ANPp}
    0 & = & i[i[\ldots i[A,P_{\mu_{N}}]\ldots,P_{\mu_{2}}],
            P_{\mu_{1}}P^{\mu_{1}}](p)  \\
      & = & A_{\vec{\mu}}\Ns(p)p^{\mu_{1}}=
            Ia_{\nu\vec{\mu}}p^{\nu}p^{\mu_{1}}
            +B_{\vec{\mu}}p^{\mu_{1}}.  \nonumber
  \end{eqnarray}
  It can be shown that $\{p^{\nu}p^{\mu}\}$ and $\{p^{\mu}\}$ are all
  linearly independent functions on $\Tm$ which means that eq.
  (\ref{ind}) follows from eq. (\ref{ANPp}). \qed
\end{quote}
Before stating the final result, one has to determine the form of the
generators involved. The generators of $\Pc$ can be obtained by
differentiating the explicit expression (\ref{BUfp}) for the
representation with respect to a parameter of a one-parameter subgroup
of $\Pc$. The generators of an ``internal'' symmetry transformations
are recognized by their commutativity with all elements of $\Pc$
(and their general form is determined using the methods described in
part 2). This gives
 
\noindent{\bf Lemma 2:}
\begin{itemize}\em
  \item  A generator of $\Lc$ is of the form
    \beq
      A(p)=B(p)+Ia_{\mu\nu}p^{\mu}\frac{1}{i}\derp{p_{\nu}},\;
      B_{[\al\lm][\al'\lm']}(p)=\dl_{\al\al'}B^{\al}_{\lm\lm'}(p)
    \eeq
    where
    \begin{description}
      \item $B(p)$ is a Hermitian matrix;
      \item $\{a_{\mu\nu}\}$ is a real antisymmetric matrix
        ($a_{\mu\nu}=-a_{\nu\mu}$).
    \end{description}
  \item A generator of $\cal T$ is of the form
    \beq  A(p)=Ia_{\mu}p^{\mu}  \eeq
    where $\{a_{\mu}\}$ is a real vector.
  \item A generator of an internal symmetry is of the form
    \beq
      A(p)=B,\;B_{[\al\lm][\al'\lm']}=B_{\al\al'}\dl_{\lm\lm'}.
    \eeq
\end{itemize}
Now it is possible to state and prove the final result:
 
\noindent{\bf Proposition 3:}
\begin{quote}\em
  A self adjoint element $A$ of $\A$ is a linear combination of
  generators of $\Pc$ and generators of internal symmetries.
\end{quote}
Proof:
\begin{description}
  \item[$N=0$:] From lemma 1 we obtain:
    \beq
      A(p)=A^{(N)}(p)=Ia_{\nu}p^{\nu}+B,\;
      B_{[\al\lm][\al'\lm']}=B_{\al\al'}\dl_{\lm\lm'}
    \eeq
    and this is recognized as a sum of a generator of translations
    (the first term) and a generator of an internal symmetry (the
    second term).
  \item[$N=1$:] From lemma 1 we have
    \beq
      A^{(1)}_{\mu}(p)=Ia_{\nu\mu}p^{\nu},\;a_{\nu\mu}=-a_{\mu\nu}
    \eeq
    which implies that
    \beq
      A(p)=A^{(0)}(p)+Ia_{\nu\mu}p^{\nu}\frac{1}{i}\derp{p_{\mu}}.
    \eeq
    The second term is the space part of a generator of $\Lc$.
    Subtracting this generator from $A$ one obtains a $0$-order
    (self adjoint) element of $\A$ which was already shown to satisfy
    the statement of the proposition.
  \item[$N>1$:] $a_{\nu\vec{\mu}}$ is symmetric in $\vec{\mu}$ (by eq.
    (\ref{Apol})). Combining this with eq. (\ref{ind}) one obtains:
    \begin{eqnarray*}
             a_{\nu\mu_{1}\mu_{2}\ldots}
      & = & -a_{\mu_{1}\nu\mu_{2}\ldots}
        =   -a_{\mu_{1}\mu_{2}\nu\ldots}
        =    a_{\mu_{2}\mu_{1}\nu\ldots}=  \\
      & = &  a_{\mu_{2}\nu\mu_{1}\ldots}
        =   -a_{\nu\mu_{2}\mu_{1}\ldots}
        =   -a_{\nu\mu_{1}\mu_{2}\ldots}
    \end{eqnarray*}
    so $a_{\nu\vec{\mu}}=0$ which implies that $A^{(N)}_{\vec{\mu}}=0$
    in contradiction to the fact that $A$ is of order $N$. Thus $N$ is
    either $0$ or $1$. \qed
\end{description}

\newsection{Comments and Supplements}
\label{Comm}

In this section we discuss the assumptions of the theorem proved in
this work, emphasizing the prospects for relaxing some of them.

\subsection{Other Signatures and Orbits}

The theorem was proved for signatures of the type $(r,1)$
(assumption 2) and for representations with momentum support in the
forward light cone (assumption 3). All this was needed to assure the
compactness of the little group, which implies that its irreducible
representation spaces $\{\Hc(\La)\}$ are finite-dimensional.
The finiteness of the dimension of the spaces
$\{\Hc(m)\}$ (which are direct sums of $\{\Hc(\La)\}$) is essential
to all of part 2 of the proof and also plays a key role in the
construction of the base of plane-wave states \cite{Horwitz-Pelc}.
It was used to construct, using the method of induced representations,
a space $\Fa$ in which the generators of $\Pc$ are represented. Such
a space can be constructed for any representation which can be built
by a {\em sequence} of inductions, starting with a finite-dimensional
representation. Perhaps the proof of the theorem may also be constructed
for such types of representations by applying the methods described in
chapter 3 successively for each stage of induction.

The infinite dimension of $\{\Hc(\La)\}$ may cause another complication.
In this case, the spectrum $\sa$ (of the operator $J$ used to
represent $\{\Hc(\La)\}$ as a space of functions -- see Section
\ref{Hal}) is not necessarily discrete. If it is continuous, $\Om$
(defined in appendix \ref{OneSpace})
is not a countable union of orbits so to consider
$\Om$ as a smooth separable manifold, one must include a differential
structure on $\sa$; this must be taken into account when checking the
smoothness of functions on $\Om$. If the spectrum is mixed, $\Om$ is
a union of manifolds of different dimension.

Finally, the choice of signature $(r,1)$ and momenta in the forward
light cone has also a physical significance. In this region $p^{0}$
is bounded from below (positive), thus suitable to be interpreted as
the energy. In any other case (except for the forward light-like
momenta in the case of signature $(r,1)$) the orbits are unbounded
in all directions, and therefore the canonical energy is not well
defined (Recall that the energy is distinguished from other
components of the momentum by being positive and this in an invariant
(and therefore well defined) statement only in the case of signature
$(r,1)$.).

The restriction $r\geq3$ was used twice. In lemma 9 of part 2 of the
proof, it assured the existence of a one-parameter subgroup of the
little group of a time-like momentum. It also assured that all
the projective representations of $\Pc$ are (equivalent to) true
representations (this is true for any signature $(r,s)$ with
$r+s\geq3$ \cite{Barg-ray}). This
was the main motivation in replacing $\Pc'\so$ with its covering group
$\Pc$. The problem with projective representations is that they do not
lead naturaly to a representation of the generators of the group.

\subsection{The Particle-Type Spectrum}

In part 1 of the proof it was assumed that the mass spectrum is
bounded; for \mbox{part 2}, $\Hc(m)$ must be finite-dimensional, which
implies that the number of particles with the same mass must be
finite; and for part 3 (eq. (\ref{AcomP})) the mass spectrum $\M$ must
consist of isolated points. Combining all this, one finds that the
number of particle types must be finite.

It might be possible to extend part 1 of the proof to an unbounded
mass spectrum (as suggested by Coleman and Mandula \cite{Col-Man}), so,
considering the other restrictions, the mass spectrum could be an
infinite increasing sequence, diverging to infinity. But
some modifications are needed. Recall (section 3.1) that the
requirement of a bounded spectrum was used to show the existence of a
sufficient variety of physical momenta outside the region
$\bbar{R}\cup\bbar{R\dg}$ (lemma 3). In the case of an unbounded mass
spectrum, $\bbar{R_{m}}\cup\bbar{R_{m}\dg}$ is spread over all of $\Tm$
for $\Dl p\so$ in the forward light cone, therefore the required
momenta must be looked for between the bands of
$\bbar{R_{m}}\cup\bbar{R_{m}\dg}$.
The width of the bands is determined by $\rho$ (the radius of the
support of the function $f$ used to construct $\fA$), but it also
depends on the angle of intersection of $\Tm$ and $\Tma$ (see figure
\ref{fig1}), therefore it cannot be bounded uniformly. Roughly speaking,
the width increases with the distance from the origin so if the intervals
between the masses don't increase accordingly, they will start to
overlap far enough  from the origin. With these considerations it is
possible to show:

\noindent{\bf Proposition:}
\begin{quote}\em
  If $\M=\{\ma\}$ is an increasing sequence and
  $\lim_{\al\rightarrow\infty}\frac{m_{\al+1}}{\ma}=1$
  (which means that $\{\ma\}$ increases slower than any geometric series)
  then, for time-like $\Dl p$ and for $\al$ sufficiently large, the
  (elliptic) bands overlap completely.
\end{quote}
This implies that under the assumptions of the proposition,
$\Tm\setminus\bbar{R}$ is bounded, therefore for $p'\so$ large enough,
the momenta satisfying lemma 3 don't exist, so the proof for this case can
not proceed as described in section 3.

Attempts to deal with this problem can be made in two different
approaches:
\begin{enumerate}
  \item Investigation of the conditions on $\M$ in which the existence
    of the required momenta can be assured: Using the same methods
    used to prove the above proposition one may establish conditions
    on $\M$ for which at least on half of $\Tm$ the bands occupy
    arbitrarily small portions of $\Tm$. This seems sufficient to show
    the existence of the required momenta, but the proof may be quite
    complicated technically.
  \item Investigation of conditions on $\M$ in which, for a given
    $\rho>0$ the required momenta exist for $p'\so$ in some region of
    $\Tm$ that approaches all of $\Tm$ at the limit
    $\rho\rightarrow0$. The arguments of part 1 imply now that the
    support of $\fA$ is outside this region. It remains to show what
    this says about $A$.
\end{enumerate}
 
\subsection{The Assumptions on the Scattering Amplitudes}
The regularity of the scattering amplitudes is used only in lemma 3 of
part 2 of the proof (see eq. (\ref{pos})), to state that the integrands
in the right hand side of
\begin{eqnarray*}
  \lefteqn{2\mbox{Im}(h\dg<p\sq|T|p\sq>h)=}  \\
  & & \sum_{m=0}^{\infty}\int_{\TF\m}d\mu\m(q\m)(2\pi)^{d}\dl^{d}
  (\sum_{1}\m q_{j}-\sum_{1}\sq p_{i})
  h\dg<p\sq|T|q\m>\dg<q\m|T|p\sq>h
\end{eqnarray*}
are non-negative and thus so are the integrals. If an analogous
argument for distributions can be given to show that the integrals are
non negative (in the sense of generalized functions) then it will not
be necessary to assume anything about amplitudes between states with
more than two particles.

The analyticity of $<T^{(2,2)}>$ is used many times but most of the
time only the analyticity of elastic scattering amplitudes (those
connecting states with the same types of particles) is really needed.
The only use of the full amplitude is to show the
$S^{(2,2)}$-invariance of $\F^{(2)}$ and there it is enough to assume
that it is smooth. However the distinction between elastic and not
elastic amplitudes seems rather artificial, since by performing an
internal transformation (mixing particle types) the ``new'' elastic
amplitudes are linear combinations of ``old'' inelastic amplitudes.
Such an argument might be used to show that the analyticity of the
elastic amplitudes in fact implies the analyticity (or at least the
smoothness) of $<T^{(2,2)}>$. Realizing that the ``diagonal'' of
$<T^{(2,2)}>$ consists only of elastic amplitudes, this seems somehow
related to the result that a sesquilinear form is determined by its
diagonal, \ie as a polarization of the form:
\[  (x,y)=\frac{1}{2}[(x+y,x+y)-i(x+iy,x+iy)]
         +\frac{i-1}{2}[(x,x)+(y,y)].  \]
 
\subsection{Super Symmetry}
  The proof of the theorem refers actually only to the algebra of
generators of symmetry (and not to the symmetry group), therefore
most of it can be applied also to supersymmetric generators, as
observed in \cite{Haag}. To include supersymmetry, one has to modify
slightly the definition of $\A$ (at the beginning of section 3). In
this case $\A$ is a direct sum
 \[ \A=\A_0\oplus\A_1\in L^{\times}(\Fone;\Foned) \]
and the ``even'' and ``odd'' elements are distinguished by their
action in $\Fone\times\Fone$. For $\f_1$ either purely even (bosonic)
or purely odd (fermionic), and $\f_2$ arbitrary, (4.2) is replaced by

\beq
  A\f:=(A\otimes I\pm I\otimes A)\f=
  (A\f_{1})\otimes\f_{2}\pm\f_{1}\otimes(A\f_{2}),
\eeq
 where the minus sign refers to $A$ and $\f_1$ both odd and the
plus sign to all other combinations (this form can be deduced from the 
Fock representation of the space of states, where the generators, are 
bilinears of creation and destruction operators).
Comparing to the original definition (in section 3), we observe that
$\A_0$ is the original $\A$ and therefore the theorem applies fully to
$\A_0$. As for all of $\A$, among its properties enumerated at the
beginning of section 4, only the last one needs modification:
\begin{description}
  \item[$4'$.] If $A\in\A_i$, $B\in\A_j$, ($i,j=0$ or $1$), $\Fone$ is
    invariant under $A$ and $B$ and \linebreak 
    $AB,BA\in L^{\times}(\Fone;\Foned)$ then $AB-(-1)^{ij}BA\in\A$.
\end{description}
This property was not used in Part 1, so proposition 1 holds for all
of $\A$. In particular, all the generators have the form (\ref{Apol})
and commute with the ``mass'' operator $P^2$. Property ($4'$) entered
the proof only in its second part, after Lemma 6, where it lead to the
conclusion that $\B^*$ is a Lie algebra. This obviously translates here
to the statement that $\B^*$ is a {\em graded} Lie algebra. Lemma 6
of part 2 (together with the form of $\A_0$, as given by the theorem)
was the starting point of \cite{Haag} and since it continues to hold
in general dimension, one can proceed as in \cite{Haag} to
determine the general form of $\A$.
 
\newsection{Conclusions}

In this work we investigated the generalization of the Coleman-Mandula
theorem to higher dimension. It states that the group of symmetries of
the (nontrivial, Poincar\'{e}-invariant) scattering matrix $S$ can
contain, in addition to the Poincar\'{e} symmetries, only
Poincar\'{e}-invariant symmetries (note that this does not exclude a
richer symmetry of the action). The theorem was proved for arbitrarily
higher spatial dimension and for a finite number of particle types,
all of them massive. Further generalization requires more involved
analysis and this was discussed in some detail in the last section.

To put the analysis on a firm basis, with minimal loss in clarity, we
used a rigorized version of the Dirac formalism, developed in \cite{RHS} 
and applied in \cite{Horwitz-Pelc} to scattering scenarios. 
Unlike other rigorous formulations of quantum mechanics,
in this formalism it is possible to use ``a complete set of plane-wave
states'' to decompose expressions into ``vector components'' and ``matrix
elements'' in almost the same flexibility as in the original Dirac
formalism. This work should also be seen as a demonstration of this
flexibility.

\appendix
\renewcommand{\newsection}[1]{
 \vspace{10mm} \pagebreak[3]
 \addtocounter{section}{1}
 \setcounter{equation}{0}
 \setcounter{subsection}{0}
 \setcounter{paragraph}{0}
 \setcounter{equation}{0}
 \setcounter{figure}{0}
 \setcounter{table}{0}
 \addcontentsline{toc}{section}{
  Appendix \protect\numberline{\Alph{section}}{#1}}
 \begin{flushleft}
  {\large\bf Appendix \thesection. \hspace{5mm} #1}
 \end{flushleft}
 \nopagebreak}

\newsection{Summary of Notation and Concepts}

\subsection{The Group $\Pc(r,s)$}
\label{Poinkare}

\begin{itemize}
  \item $\Pc_0\equiv\Pc_0(r,s)$ is the restricted (identity component
    of) inhomogeneous pseudo-orthogonal group of signature $(r,s)$;
  \item $\Lc_0\equiv\Oc_0(r,s)$ is the homogeneous part of $\Pc_0(r,s)$;
  \item $\Tc\equiv\Tc_{r+s}$ is the translation group in ${\Rb}^{r+s}$;
  \item $\Pc$ and $\Lc$ are the universal covering groups of $\Pc_0$ and
    $\Lc_0$ respectively.
\end{itemize}
$\Pc$ ($\Pc_0$) is the semi-direct product of $\Tc$ and $\Lc$ ($\Lc_0$).
The theorem is proved for $s=1$ and $r\geq3$ (assumption 2) and these are
the values assumed also in the Appendices. For these values
$\Lc_0=\Lc/\Zb_2$.

\subsection{The Momentum Space}
\label{momentum}

\begin{itemize}
  \item $\T$, ``the momentum space'': the dual of the translation
    group $\Tc$;
  \item $\Tm$, ``A mass shell'': an orbit of $\Lc$ in $\T$; assumed
    to be in the forward light cone (assumption 3 of the theorem), so the
    elements $\{p\}$ of $\Tm$ are characterized by the ``mass''
    $m=\sqrt{p_{\mu}p^{\mu}}$;
  \item $\mm$ : the $\Lc$-invariant non-trivial Radon measure on $\Tm$
    (unique up to a multiplicative constant); it is non-degenerate, in
    the sense that it does not vanish on open sets;
  \item $\M$, ``the one-particle mass spectrum'': the set of masses
    of the particles of the system; assumption 4 of the theorem implies
    that it is a finite set;
  \item $\TF:=\bigcup_{m\in\M}\Tm$ : the physical region in $\T$ for
    one-particle states.
\end{itemize}

\subsection{The Space $\Ha$ of $\al$-states}
\label{Hal}

A particle type $\al$ is identified with an irreducible representation
$\Ua$ of $\Pc$ in the space of one-particle states.
$\Ha$ -- ``the $\al$-states space'' is the representation space
of $\Ua$ (the space of all possible states in which there is one
particle and it is of type $\al$).

After a spectral decomposition of $\Ha$, one gets:
\beq  \Ha=\bigoplus_{\lm\in\sa}\Lsq_{\mma}(\Tma)  \eeq
where
\begin{itemize}
  \item $\Tma$, ``the $\ma$-mass shell'' is the
    spectrum of the momentum operator $P$ in $\Ha$;
  \item $\sa$ is the spectrum of the operator $J\in\UP$ 
    (where $\UP$ denotes the universal enveloping algebra of the Lie
    algebra of $\Pc$) which supplements $P$ to a complete system of commuting
    observables in $\Ha$; assumption 3 of the theorem implies that $\sa$
    is a finite set;
\end{itemize}
so, for each $f\in\Ha$
\begin{eqnarray}
  {}[\Ua(\Pm)f](p,\lm)    & = & p^{\mu}f(p,\lm)  \\
  {}[\Ua(J_{i})f](p,\lm) & = & \lm_{i}f(p,\lm)
\end{eqnarray}

\subsection{The One-Particle Space $\Hone$}
\label{OneSpace}

\beq  \Hone=\bigoplus_{\al\in I}\Ha=\Lsq_{\mu}(\Om)  \eeq
where
\begin{itemize}
  \item $I$ is the set of particle types (a finite set according to
    assumption 4 of the theorem);
  \item $\Om:=\{(p,\lm,\al)|\al\in I,\lm\in\sa,p\in\Tma\}$ (it is an
    $r$-dimensional separable smooth manifold);
  \item $\mu$ is the (non-degenerate Radon) measure on $\Om$ defined
    by
    \beq
      \int_{\Om}d\mu(p,\lm,\al)\ldots=
      \sum_{\al\in I}\sum_{\lm\in\sa}\int_{\Tma}d\mma(p)\ldots
    \eeq
\end{itemize}
Considering an element of $\Hone$ as a vector-valued function on the
momentum space $\T$, we write:
\beq  \Hone=\bigoplus_{m\in\M}\Lsq_{\mm}(\Tm,\Hc(m))  \eeq
where
\begin{itemize}
  \item $I(m):=\{\al\in I|\ma=m\}$ is the set of particle types with
    mass $m$;
  \item $\Om(m):=\{[\al\lm]|\al\in I(m),\lm\in\sa\}$ (this is a finite
    set according to assumptions 3 and 4 of the theorem);
  \item $N(m)$ is the number of elements in $\Om(m)$;
  \item $\Hc(m)={\bf C}^{N(m)}$ (the $N(m)$-dimensional complex
    Hilbert space).
\end{itemize}
When $m=\sqrt{p_{\mu}p^{\mu}}$, $I(p),\Om(p),N(p),\Hc(p)$ stand for
$I(m)$ etc.\ .

\noindent With this approach, $\mu$ can be seen as a measure on $\TF$:
\beq
  \int_{\Om}d\mu(p,\lm,\al)\ldots=
  \int_{\TF}d\mu(p)\sum_{[\al\lm]\in\Om(p)}\ldots
\eeq
where
\beq
  \int_{\TF}d\mu(p)\ldots:=\sum_{m\in\M}\int_{\Tm}d\mm(p)\ldots\;.
\eeq

\subsection{The Full Space of States}

The $n$-particle space $\Hns$ is a closed subspace of
\beq
  \Hn=\bbar{\bigotimes_{1}\n\Lsq_{\mu}(\Om)}
     =\Lsq_{\mu\n}(\Om\n)
\eeq
(the bar denotes closure in $\Lsq_{\mu\n}(\Om\n)$ ) where
\begin{itemize}
  \item $\Om\n:=\Om\times\cdots\times\Om$ ($n$ factors)

    (this is a separable smooth manifold);
  \item $\mu\n$ is the ($\Lc$-invariant non degenerate Radon) measure
    defined by
    \beq
      \int_{\Om\n}d\mu\n(p\n,\lm\n,\al\n)\ldots:=
        \int_{\Om}d\mu(p_{1},\lm_{1},\al_{1})\ldots
        \int_{\Om}d\mu(p_{n},\lm_{n},\al_{n})\ldots
    \eeq
    or, in vector notation:
    \beq
      \int_{\TF\n}d\mu\n(p\n)\ldots:=
        \int_{\TF}d\mu(p_{1})\ldots\int_{\TF}d\mu(p_{n})\ldots
    \eeq
    where $\TF\n:=\TF\times\cdots\times\TF$ ($n$ factors).
\end{itemize}
The elements of  $\Hns$ are those elements of $\Hn$ which have the
right symmetry properties with respect to exchange of identical
particles.

\noindent The space $\Hs$ of all physical states is
\beq  \Hs=\bigoplus_{n=0}^{\infty}\Hns  \eeq
and thus it is a closed subspace of
\beq  \Hc=\bigoplus_{n=0}^{\infty}\Hn.  \eeq
For $f\in\Hc$ we write
\beq  f=\sum_{n}f\n,\;f\n\in\Hn.  \eeq
For $f\n\in\bigotimes_{1}\n\Hone$ we write
\beq  f\n=\bigotimes_{i}f_{i}\n,\;f_{i}\n\in\Hone.  \eeq
For $\f\in\Hn,p\n\in\Om\n$, $\f(p\n)$ is a vector in
\beq  \Hc(p\n):=\bigotimes_{1}\n\Hc(p_{i}).  \eeq

\subsection{Gel'fand Triples}
\label{Gelfand}

The Gel'fand triples (see section 2.1) for the various spaces defined
above are obtained by defining:
\beq\begin{array}{lllll}
  \mbox{for }\Ha   & : & \Fa   & := & \bigoplus_{\lm\in\sa}\D(\Tma) \\
  \mbox{for }\Hone & : & \Fone & := & \D(\Om)=\bigoplus_{\al\in I}\Fa \\
  \mbox{for }\Hn   & : & \Fn   & := & \D(\Om\n) \\
  \mbox{for }\Hns  & : & \Fns  & := & \Fn\cap\Hns  \\
  \mbox{for }\Hc   & : & \F    & := & \bigoplus_{n=0}^{\infty}\Fn  \\
  \mbox{for }\Hs   & : & \Fs   & := &
                               \F\cap\Hs=\bigoplus_{n=0}^{\infty}\Fns
\end{array}\eeq
(The direct sums are as defined for locally convex spaces and are the
sets of {\em finite} sums of elements.)

The commuting set of observables includes the components of the
momentum operator $P^{\mu}$ therefore the basis elements are
``plane waves'':
\beq  \{<p\n|\;|n=0,1,\ldots,p\n\in\TF\n\}  \eeq
and the expression that plays the role of the identity operator is
\beq  I:=\sum_{n=0}^{\infty}\int_{\TF\n}d\mu\n(p\n)|p\n><p\n|  \eeq
The matrix elements $<q\m|A|p\n>$ of an operator
$A\in L^{\times}(\F;\F')$ are the generalized ``values'' of the kernels
$<A\mn>\in\D'(\Om\m\times\Om\n)$ satisfying
\beq\label{BpsAf}
  (\psi|A|\f)=\sum_{m,n=0}^{\infty}\int_{\TF\m}d\mu\m(q\m)
    \int_{\TF\n}d\mu\n(p\n)(\psi|q\m><q\m|A|p\n><p\n|\f).
\eeq
Since $\f(p\n)$ is a vector in $\Hc(p\n)\equiv\otimes_1^n\Hc(p_i)$, eq.
(\ref{BpsAf}) implies the following interpretation:
\begin{quote}
  $<p\n|\f)$ (and therefore also $<p\n|$) is a column vector of
  dimension dim$\Hc(p\n)$;

  $(\psi|q\m>$ (and therefore also $|q\m>$) is a row vector;

  $<q\m|A|p\n>$ is a matrix operator from $\Hc(p\n)$ to $\Hc(q\m)$.
\end{quote}

\subsection{The Representation $U$ of $\Pc$}

\subsubsection{The Irreducible Representation $\Ua$ in $\Ha$}
\label{RepAl}

We denote
\beq  \psm:=(m,0,\ldots,0)\mbox{ (``the rest frame'')}  \eeq
and choose a smooth function $p\mapsto\Lmp$ from $\Tm$ to $\Lc$
obeying:
\begin{enumerate}
  \item $\Lm_{\psm}=1$ (the unit matrix)
  \item $\Lmp\psm=p,\;\forall p\in\Tm$
\end{enumerate}
Now $\Ua$ is (for all $(\Lm,a)\in\Pc,f\in\Ha,p\in\Tma$)
\beq\label{BUfp}
  [\Ua(\Lm,a)f](p)=e^{ip\cdot a}\La(\Dl(\Lm,p))f(\Lm\inv p)
\eeq
where $\Dl(\Lm,p)$ is defined by
\beq\label{BDLp}
  \Dl(\Lm,p)=\Lmp\inv\Lm\Lm_{\Lm\inv p}\in\Lc(\pma)
\eeq
and $\La$ is a continuous unitary irreducible matrix representation
of the {\em little group} $\Lc(\pma)$, which in this case is the universal
covering group of $\Oc(r)$.

\subsubsection{The Representation in $\Hc$}

The representation in $\Hone$ is $\Uone=\bigoplus_{\al\in I}\Ua$ so
(for all $(\Lm,a)\in\Pc,f\in\Hone,p\in\TF$)
\beq  [\Uone(\Lm,a)f](p)=e^{ip\cdot a}\Lp(\Dl(\Lm,p))f(\Lm\inv p)  \eeq
where $\Lp=\bigoplus_{\al\in I(p)}\La$ is the (reducible) unitary
representation of $\Lc(\psm)$ in $\Hc(m)$ \linebreak
($m=\sqrt{p_{\mu}p^{\mu}}$).

The representation $U\ns$ in $\bigotimes_{1}\n\Hone$ is
\beq
  U\ns(f_{1}\otimes\cdots\otimes f_{n})=
  (\Uone f_{1})\otimes\cdots\otimes(\Uone f_{n}),\;
  \forall f_{i}\in\Hone.
\eeq
This is extended to $\Hn$ by continuity and to $\Hc$ by linearity.

\subsubsection{The Matrix Elements of $\Uone(\Pc)$}
The action of $\Uone$ on the base vectors is
\beq  \Uone(\Lm,a)|p>=|\Lm p>e^{ia\cdot\Lm p}\Lp(\Dl(\Lm,\Lm p))  \eeq
and with components:
\beq
  \Uone(\Lm,a)|p,\lm,\al>=
  e^{ia\cdot\Lm p}\La(\Dl(\Lm,\Lm p))_{\lm'\lm}|\Lm p,\lm',\al>.
\eeq
The matrix elements of $\Uone$ are
\beq
  <p'|\Uone(\Lm,a)|p>=
  e^{ia\cdot\Lm p}\Lp(\Dl(\Lm,\Lm p))\dl_{\mu}(p'-\Lm p)
\eeq
and with components:
\beq
  <p',\lm',\al'|\Uone(\Lm,a)|p,\lm,\al>=e^{ia\cdot\Lm p}
  \dl_{\al'\al}\La(\Dl(\Lm,\Lm p))_{\lm'\lm}\dl_{\mu}(p'-\Lm p).
\eeq

\subsection{Generators of Symmetry}
\label{Gen}

A one-parameter symmetry group $g(t)$ of $S$ is represented in $\Hn$
by a unitary representation $\Un$. The {\em generator $A_g^{(n)}$ of
$\Un(g(t))$} is defined by
\beq
  (\psi|\Ag^{(n)}\f\!>:=\frac{1}{i}\der{t}(\psi,\Un(g(t))\f)|_{t=0}
  \hsc \forall\f,\psi\in\Fn.
\eeq
$A_g\equiv A_g^{1}$ is assumed (assumption 5 of the theorem) to be an
element of $L^{\times}(\Fone;\Foned)$ (a continuous antilinear map from
$\Fone$ to $\Foned$) and it is self adjoint. If $g(t)$ is a subgroup of
$\Pc$, $A_g$ is, by construction of $\Fone$, a continuous operator in
$\Fone$ (and therefore certainly satisfies assumption 5).

\noindent Between elements of $\bigoplus_{1}\n\Fone$:
\beq\label{Agn}
  \Ag\ns=(\Ag\otimes I\otimes\cdots\otimes I)
        +(I\otimes\Ag\otimes\cdots\otimes I)
        +\cdots+(I\otimes I\otimes\cdots\otimes \Ag).
\eeq
Being a generator of symmetry, $A_g$ satisfies
(when $S^{*}\psi,S\f\in\F$)
\beq  (S^{*}\psi|\Ag\f\!>=<\!\Ag\psi|S\f).  \eeq
In particular, if $S$ and $\Ag$ are operators in $\F$ then
\beq\label{com}  [\Ag,S]=0\mbox{ in }\F.  \eeq
Also, if $\Ag$ is a {\em continuous operator in} $\F$ (\eg a generator
of $\Pc$) then (\ref{com}) holds, with the commutators defined to be
\beq  [\Ag,S]=\Ag'S-S\Ag  \eeq
where $\Ag'$ is the dual of $\Ag$ and $S$ is considered as an operator
from $\F$ to $\F'$.

\subsection{Scattering Amplitudes}
\label{scatter}

The S-matrix $S$, being unitary, can be identified as an element of
$L^{\times}(\F;\F')$. As such, it has a corresponding kernel $<\!S\!>$
(more precisely -- kernels $<\!S\mn\!>\in\D(\Om\m\times\Om\n)$ ).

The translation invariance of $S$ implies that $<\!S\!>$ has the
following form
\beq\label{SIT}
  <\!S\mn\!>=<\!I\mn\!>
    -i(2\pi)^{d}\dl^{d}(\sum_{1}\m q_{j}-\sum_{1}\n p_{i})<\!T\mn\!>
\eeq
where $d=r+1$ is the dimension of the momentum space, $-i(2\pi)^{d}$
is a conventional normalization factor and $<\!T\!>$ is a generalized
function on the submanifold of $\Om\m\times\Om\n$ defined by the
constraint
\beq  \sum_{1}\m q_{j}-\sum_{1}\n p_{i}=0.  \eeq
(this is the precise formulation of energy-momentum conservation).
The values of $<\!T\!>$ are called ``scattering amplitudes''.

Since $S$ is $\Lc$-invariant, $<\!T\!>$ depends only on $\Lc$-invariant
quantities. In particular, its dependence on the momenta is only
through $\Lc$-invariant functions of the momenta.

The unitarity of $S$ leads to

\noindent{\bf The Optical Theorem:}
\begin{eqnarray}
  \lefteqn{<\!p\n|T|p\n\!>-<\!p\n|T|p\n\!>\dg=}  \\
  & & i\sum_{m=0}^{\infty}\int_{\TF\m}d\mu\m(q\m)(2\pi)^{d}\dl^{d}
      (\sum_{1}\m q_{j}-\sum_{1}\n p_{i})
      <\!p\n|T|q\m\!>\dg<\!q\m|T|p\n\!>   \nonumber
\end{eqnarray}
where ``$\dg$'' denotes the Hermitian conjugation of (matrix) maps
between the spaces $\Hc(p^n)$.

\newsection{Two-Particle Scattering}

\subsection{The $\Lc$-Invariant Variables}

The scattering amplitudes depend on the momenta through $\Lc$-invariant
functions. We show here that for the scattering between two-particle states
\beq\label{pppp} (p_{1},p_{2})\longrightarrow(p'_{1},p'_{2}) \eeq
(where the momenta are, according to assumption 3 of the theorem, in the
forward light cone) these functions can be chosen to be the traditional
\beq s:=(p_{1}+p_{2})\sq  \eeq
and
\beq  t:=(p'_{1}-p_{1})\sq.  \eeq
Explicitly we show that $s$ and $t$ determine the momenta in the process
(\ref{pppp}) up to an $\Lc$-transformation (represented by an $\Lc\so$
transformation).

We start with the identity
\beq
  (p_{1}+p_{2})\sq=
    m_{1}\sq+m_{2}\sq+2E_{1}E_{2}-2\pv_{1}\cdot\pv_{2}.
\eeq
Since $\pv_{1}\cdot\pv_{2}\leq|\pv_{1}||\pv_{2}|<E_{1}E_{2}$, we have
$(p_{1}+p_{2})\sq>0$, which means that $p_{1}+p_{2}$ is time-like and
can be transformed, by an $\Lc\so$-transformation to the ``rest frame'',
where $\pv_{1}+\pv_{2}=0$. Then one can perform a rotation (an element
of ${\cal O}_{0}(r)$, which is an $\Lc\so$-transformation), to align
$\pv_{1}$ along the $e_{1}$ axis, obtaining $\pv_{1}=(p,0,\ldots,0)$.
The equality $\pv_{1}+\pv_{2}=0$ is not affected by this transformation,
so $\pv_{2}=(-p,0,\ldots,0)$ and the conclusion is that the initial
state is characterized by one variable $p$, that can be expressed by
the invariant variable
\beq
  s=(p_{1}+p_{2})\sq=m_{1}\sq+m_{2}\sq+
    2\sqrt{(m_{1}\sq+p\sq)(m_{2}\sq+p\sq)}-2p\sq.
\eeq
(if $m_{1}=m_{2}=m$, this simplifies to $s=4(m\sq+p\sq)$.)

Momentum conservation gives
\beq  \pv_{1}\,\!'+\pv_{2}\,\!'=\pv_{1}+\pv_{2}=0.  \eeq
Substituting $(p')\sq\:=(\pv_{1}\,\!')\sq=(\pv_{2}\,\!')\sq$ in the energy
conservation equation, one obtains:
\beq
  \sqrt{((m'_{1})\sq+(p')\sq)}+\sqrt{((m'_{2})\sq+(p')\sq)}=
  \sqrt{(m_{1}\sq+p\sq)}+\sqrt{(m_{2}\sq+p\sq)}
\eeq
and this has at most one solution for $(p')\sq$. (Such solution exists
iff $\sqrt{s}\geq m'_{1}=m'_{2}$; for an elastic scattering
($m'_{i}=m_{i}$) this is $p'=p$.)

It is left to determine the direction of $\pv_{1}\,\!'$. Rotating around
$\pv_{1}$, one can bring $\pv_{1}\,\!'$ to the $(e_{1}e_{2}$) plane (notice
that such rotation doesn't affect $\pv_{1},\pv_{2}$ and $p_{1}+p_{2}$).
So, to characterize the final state, it is enough to give the angle
between $\pv_{1}$ and $\pv_{1}\,\!'$, and this can be expressed by the
invariant variable $t=(p_{1}-p'_{1})\sq$.

\subsection{The $S^{(2,2)}$-Invariance of $\F^{(2)}$}

In this subsection we show that $\F^{(2)}$ is $S^{(2,2)}$-invariant
and comment about the $S$-invariance of $\F$.

First consider the support of $(S\f)\m$ for an arbitrary $\f\in\F$.
Since $S$ conserves energy and momentum, $(p'_{1},\ldots,p'_{m})$ can
be in the momentum support of $(S\f)\m$ only if there exists an integer
$n$ and $(p_{1},\ldots,p_{n})$ in the momentum support of $\f\n$ that
satisfies $\sum_{1}\m p'_{j}=\sum_{1}\n p_{i}$. For each $n$, the
momentum support
$\{(p_{1},\ldots,p_{n})|\f\n(p_{1},\ldots,p_{n})\neq0\}$ of
$\f$ is bounded in $\TF\n$ (where $\TF$ is the ``physical region'' in
$\T$, see section \ref{momentum}) and thus
$\{\sum_{1}\n p_{i}|\f\n(p_{1},\ldots p_{n})\neq0\}$ is bounded in the
momentum space $\T$.  $\f$ has a finite number of non-vanishing
components $\{\f\n\}$, thus
$\{\sum_{1}\n p_{i}|n=0,1,\ldots,\;\f\n(p\n)\neq0\}$ is also bounded
(being a finite union of bounded sets). So we conclude that for
$(p'_{1},\ldots,p'_{n})$ in the momentum support of $(S\f)\m$,
$\sum_{1}\m p'_{j}$ is restricted to a bounded set in $\T$, and in
particular, $\sum_{1}\m E'_{j}$ is bounded. But $\{p'_{j}\}$ are in
the forward light cone, so $E'_{j}>0,\;\forall j$, and thus each $E'_{j}$
is bounded separately. $\pv_{j}\,\!'$ is bounded by $E'_{j}$ so $p_{j}$ is
bounded in $\T$. Therefore the momentum support of $(S\f)\m$ is bounded in
$\TF\m$ and thus compact; the finite dimension of
$\Hc(p'_{1},\ldots,p'_{m})$ then implies that $(S\f)\m$ has compact
support in $\Om\m$.

The smoothness of $S^{(2,2)}\f\equiv(S\f)\sq$, for $\f\in\F^{(2)}$
follows from the analyticity of the scattering amplitudes $<T^{(2,2)}>$,
so we can conclude that $S^{(2,2)}\f$ is in $\F^{(2)}$.

If, in addition, {\em all} the scattering amplitudes are
smooth, one can show that $\F$ is $S$-invariant. For $\f\in\F$, the
smoothness of the scattering amplitudes implies that $(S\f)\m$ is
smooth and the above analysis shows that supp$(S\f)\m$ is compact,
which means that $(S\f)\m\in\F\ms$. It is left to
show that $(S\f)\m=0$ for $m$ large enough. This follows from
\beq\label{Ebound}
  \sum_{1}\n E_{i}=\sum_{1}\m E'_{j}\geq\sum_{1}\m m_{\al_{j}}
    \geq m\inf(\M)
\eeq
since Inf($\M$) is positive (according to assumption 3 and 4 of the
theorem), so for $m$ large enough, (\ref{Ebound}) cannot be satisfied.

\newpage
Captions

Figure \ref{fig1}: The physical region $\TF$

Figure \ref{fig2}: A cross section of the momentum space at the plane 
$(p_{x},E)$

Figure \ref{fig3}: A projection of $\Tm$ on the $\pv$-hyperplane


\begin{thebibliography}{99}
\bibitem{Col-Man}
  S. Coleman and J. Mandula, Phys.\ Rev.\ {\bf 159} (1967) 1251
\bibitem{Vol-Aku}
  D. V. Volkov and V. P. Akulov, Phys.\ Lett.\ {\bf B46} (1973) 109.
\bibitem{Wes-Zum}
  J. Wess and B. Zumino, Nuc.\ Phys.\ {\bf B70} (1974) 39.
\bibitem{Haag}
  R. Haag, J. T. \L{}opusza\'{n}ski, and M. Sohnius, Nuc.\ Phys.\
    {\bf B88} (1975) 257.
\bibitem{Time}
  L. P. Horwitz, and C. Piron, Helv.\ Phys.\ Acta {\bf 46} (1973) 316;

  L. P. Horwitz, R. I. Arshansky and A. C. Elitzur, Found.\ of Phys.\
    {\bf 18} (1988) 1159;

  L. P. Horwitz, Found.\ of Phys.\ {bf 22} (1992) 421.
\bibitem{mult-tau}
  I. T. Todorov, Phys.\ Rev.\ {\bf D3} 2351 (1971).

  H. W. Crater and P. Van Alstine, Nuc.\ Phys.\ B (Proc. Suppl.)
    {\bf 6} (1989) 271-274.

  J. Bijtebier, Nuc.\ Phys.\ B (Proc. Suppl.) {\bf 6} (1989) 278-280.
\bibitem{ext-sym}
  This by itself is not yet a contradiction to the theorem since this 
  extended symmetry is a symmetry of the equations of motion and not 
  necessarily of the $S$ matrix.
\bibitem{Witten}
  For example, from the structure of the GLA in a system without
  scalar generators, as determined by the theorem of Haag
  {\em et al.\ }one gets that in the rest frame of the system
  $\{Q_{i},Q_{j}\}=\delta_{ij}H$ and $[Q_{i},H]=0$
  where $H$ is the hamiltonian and ${Q_{i}}$ are the odd generators.
  this is the starting point of the Witten model (E. Witten, 
  Nuc.\ Phys.\ {\bf B185} (1981) 513) which
  is a simple model in supersymmetric quantum mechanics. It would be
  desirable  to have an analogue of such a model in relativistic
  quantum mechanics, as described in the introduction. This could be 
  important in the development of a theory for a particle with spin.
\bibitem{Am-Re}
  J. M. Amig\'{o} and H. Reeh, Fortschr. Phys. {\bf 36} (1988) 929; 
  and references quoted there.
\bibitem{Bu-Lo-Ra}
  P. Buchholz, J. T. \L{}opusza\'{n}ski, and Sz. Rabsztyn, 
  Nucl. Phys. {\bf B263} (1986) 155.
\bibitem{Garber}
  W. D. Garber, J. Math. Phys. {\bf 24} (1983) 1256.
\bibitem{Strube}
  D. Strube, J. Phys. {\bf A18} (1985) 2603; 
  J. Math. phys {\bf 31} (1990) 2244; {\bf 33} (1992) 808.
\bibitem{Dirac}
  P. A. M. Dirac, {\em The Principles of quantum Mechanics},
    Clarendon Press, Oxford, England, 1930 (1st eddition), 1947
    (3rd eddition).
\bibitem{RHS}
  J. E. Roberts, J.\ Math.\ Phys.\ {\bf 7} (1966) 1097; 
  Commun.\ Math.\ Phys.\ {\bf 3} (1966) 98.

  A. B\"{o}hm, {\em The Rigged Hilbert Space in Quantum Physics}, 
  in {\em Boulder Lectures in Theoretical Physics}, A.O.Barut 
  (ed.), vol. 9A (1966)

  J.-P. Antoine, J. Math.\ Phys.\ {\bf 10} (1969) 53, 2276.

\bibitem{Horwitz-Pelc}
  L.P. Horwitz and O. Pelc, {\em Construction of a Complete Set of
  States in Relativistic Scattering Theory}, RI-1-96, TAUP 2312-95 preprint.
\bibitem{Gel'fand}
  I. M. Gel'fand, G. E. Shilov, and N. Ya. Vilenkin, {\em
    Obobshchennye Funktsii i Deistviya Nad Nimi}, Gosudarstvennoe
    Izdatel'stvo Fiziko-Matematicheskoi Literatury, Moskow, 1958-1960
    Vols.\ I-V\@. English translation: {\em Generalized Functions},
    Academic Press Inc., New York, 1964.
\bibitem{Barg-ray}
  V. Bargmann, Ann.\ Math.\ {\bf 59} (1954) 1.
\end{thebibliography}
\end{document}